# Annealing Free, Clean Graphene Transfer using Alternative Polymer Scaffolds

Joshua D. Wood[1,2,3†], Gregory P. Doidge[1,2,3], Enrique A. Carrion[1,3], Justin C. Koepke[1,2], Joshua A. Kaitz[2,4], Isha Datye[1,2,3], Ashkan Behnam[1,3], Jayan Hewaparakrama[1,3], Basil Aruin[1,2], Yaofeng Chen[1,2], Hefei Dong[2,4], Richard T. Haasch[5], Joseph W. Lyding[1,2*], and Eric Pop[6*]

[1]*Dept. of Electrical & Computer Eng., Univ. Illinois at Urbana-Champaign, Urbana, IL 61801*
[2]*Beckman Institute, Univ. Illinois at Urbana-Champaign, Urbana, IL 61801*
[3]*Micro and Nanotechnology Lab, Univ. Illinois at Urbana-Champaign, Urbana, IL 61801*
[4]*Dept. of Chemistry, Univ. Illinois at Urbana-Champaign, Urbana, IL 61801*
[5]*Materials Research Laboratory, Univ. Illinois at Urbana-Champaign, Urbana, IL 61801*
[6]*Dept. of Electrical Engineering, Stanford Univ., Stanford, CA 94305*

We examine the transfer of graphene grown by chemical vapor deposition (CVD) with polymer scaffolds of poly(methyl methacrylate) (PMMA), poly(lactic acid) (PLA), poly(phthalaldehyde) (PPA), and poly(bisphenol A carbonate) (PC). We find that optimally reactive PC scaffolds provide the cleanest graphene transfers without any annealing, after extensive comparison with optical microscopy, X-ray photoelectron spectroscopy, atomic force microscopy, and scanning tunneling microscopy. Comparatively, films transferred with PLA, PPA, and PMMA have a two-fold higher roughness and a five-fold higher chemical doping. Using PC scaffolds, we demonstrate the clean transfer of CVD multilayer graphene, fluorinated graphene, and hexagonal boron nitride. Our annealing free, PC transfers enable the use of atomically-clean nanomaterials in biomolecule encapsulation and flexible electronic applications.

* Correspondence should be addressed to lyding@illinois.edu and epop@stanford.edu.

†Present affiliation: Northwestern University, Department of Materials Science and Engineering, Evanston, IL 60208 USA

The outstanding electronic,[1] thermal,[2] and mechanical[3] properties of graphene have engendered considerable research in this two-dimensional, atomically thin carbon sheet. Initial studies used graphite exfoliation[1, 4, 5] to isolate graphene, producing high quality but relatively small samples (e.g. <40 μm). Scalability concerns were addressed partially by the chemical vapor deposition (CVD) of graphene on transition metals like Ni,[6] Ni–Cu alloy,[7] and Cu.[8-11] CVD of graphene on Cu has proven the most fruitful platform for large-area graphene growth, as the low carbon solubility promotes monolayer growth.[8] Nevertheless, most applications using CVD-grown graphene require that the films be transferred to insulating substrates. The predominant graphene transfer approach is by using a poly(methyl methacrylate) (PMMA) scaffold.[12-17] In this method, the PMMA polymer coats the graphene, supporting it during Cu removal, underside contaminant cleaning, and placement on its destination substrate.[18, 19]

However, PMMA removal from graphene after film transfer has proven challenging.[15] Approaches to remove it by high-temperature Ar/H₂ forming gas annealing,[14, 20, 21] O₂ based annealing,[15, 22, 23] and in situ annealing[16, 24, 25] have been marginally successful in removing PMMA without affecting the graphene. Furthermore, these processes are all at high-temperature, excluding graphene applications with low thermal budgets, including uses in flexible electronics and biomolecule encapsulation. Another process separated the graphene from the PMMA support by an Au interfacial layer,[26] but that process is subject to effective interfacial Au-graphene wetting. Recent transfer results using thermal release tape (TRT),[27-29] poly(bisphenol A carbonate) (PC),[30, 31] and sacrificial polymer release layers[26] required elevated temperature (over 100°C) during transfer and differed considerably in terms of surface contamination and graphene area coverage. To exploit the intrinsic properties of large-area graphene, a room temperature transfer process that comes off more cleanly than the established methods is needed.

In this study, we compare the transfer of graphene with the conventional PMMA polymer scaffold with alternative poly(lactic acid) (PLA), poly(phthalaldehyde) (PPA), PC, and bilayer PMMA/PC scaffolds. We choose both PLA and PPA as scaffolds as they can supposedly be removed by modest heating or acid exposure. Further, we choose PC from its heightened reactivity as a condensation polymer and its former use[30, 31] in small-area graphene transfer. We find that PC scaffolds can be fully removed off the graphene by room temperature dissolution in chloroform. Contrasted against previous work, our process produces large-area graphene transfers, highlights the amount of polymer contamination clearly, and examines the fundamental chemistries involved in transfer polymer dissolution. Additionally, PC-transferred graphene samples possess cleanliness to the atomic scale, as compared to the room temperature removal of PMMA, PLA, and PPA. The PC transfer process is general, allowing us to cleanly layer two-dimensional materials like graphene, CVD fluorinated graphene (FG), and CVD hexagonal boron nitride (h-BN).

We grow graphene by CVD on Cu using previously established recipes.[8, 11, 24] More details are given in the Supporting Information (SI). We spin coat the PMMA,[21, 24, 25] PLA,[32] PPA,[33] PC,[30, 31] and PMMA/PC scaffolds using processes also outlined in the SI. After fabricating these scaffolds, we etch the Cu away and mainly transfer to thermally-grown 90 nm $SiO_2$ on Si ($SiO_2$) substrates. Water is trapped in this process (see SM1, SI), but other solvents can be used (Fig. S1, SI). Further, we transfer to other nanomaterial layers (e.g. graphene and h-BN) on $SiO_2$ and to mica substrates.[24] Room temperature removal of the scaffolds takes place in a chloroform bath for at least 1 hr; we have considered other solvents for PMMA removal and found them ineffective (see SI). We also use thermal release tape (TRT), photoresist (AZ5214 PR), and aromatic poly(aniline) (PANI) based transfers, all revealed in the SI. The former transfer method results in many holes in our samples, and the latter methods give brittle films with substantial residue. Therefore, we do not employ these transfer strategies.

Figure 1a shows our transfer process for polymer-based graphene scaffolds. The schematic demonstrates that, even with the typical Ar/$H_2$ anneal, polymer contaminants remain on the graphene after transfer. Without this anneal, the polymer contamination level on the graphene is considerably worse. Still, we are interested in a polymer removal process that takes place at room temperature, and thus we will focus on non-annealed samples for the majority of the following work. In Fig. 1b, we give the chemical formulae for aliphatic PMMA, aliphatic PLA, aromatic PPA, and aromatic PC. Figures 1c-g show optical images of graphene transferred with different polymer scaffolds on $SiO_2$, with the scaffolds dissolved in chloroform at room temperature. We give the polymer scaffolds' thicknesses in the supporting information, as determined by profilometry. The polymer repeating unit is shown in Fig. 1b for each of the different scaffolds. In Fig. 1c, the PMMA-transferred film is continuous, with no contamination optically evident. Conversely, in Fig. 1d, the PLA-transferred film is discontinuous, with folded and contaminated edges. This suggests that the PLA transfer scaffold is less elastic, less robust, and interacts with the graphene more strongly than PMMA. Figure 1e shows the PPA-transferred graphene film; it too is discontinuous and contaminated like the PLA-transferred graphene. In contrast, the PC and PMMA/PC bilayer transferred graphene films in Figs. 1f and 1g, respectively, appear continuous and uncontaminated, like the PMMA-transferred graphene films. Selected area electron diffraction (SAED) measurements on the PMMA/PC sample—shown inset in Fig. 1g—reveal monolayer graphene domains within a single CVD graphene grain. The single set of diffraction spots suggests that turbostratic ordering from transfer-induced folds[25, 34] is non-existent.

We can assess the chemistries present on the graphene surface after polymer dissolution by means of X-ray photoelectron spectroscopy (XPS). Within Fig. 1h, we show offset C 1s photoelectron spectra for PLA, PMMA and PC transferred graphene films, all of which were from the same graphene growth (curves offset for clarity). Several reports analyzing the thermal decomposition of PMMA on graphene via XPS[12, 16, 35] fit sub-peaks for the C–C backbone, the –CH$_3$ subgroup, oxygenated (ester and ether) functionals, and others in the PMMA repeating unit. It is challenging to discriminate conclusively amongst these sub-groups and the innate functionals introduced by graphene CVD growth and ambient exposure. We consider the sp$^2$ C, sp$^3$ C (introduced by –CH$_3$ and others), carboxyl C–O, carbonyl C=O, oxygenated aryl,[36] and carbonate CO$_3$ subpeaks in our PLA-, PMMA-, and PC-transferred graphene films (see SI). We find that the amount of residual functionals relative to graphene (sp$^2$ sub-peak with sp$^3$ contributions removed) is 28.4%, 11.2%, and 2.1% for PLA-, PMMA-, and PC- transferred graphene, respectively. At a typical PMMA doping concentration of $p \sim 1\times10^{12}$ cm$^{-2}$,[16] these amounts correspond to concentrations of $3\times10^{12}$ cm$^{-2}$ and $2\times10^{11}$ cm$^{-2}$ for PLA- and PC-transferred graphene. Doping levels of $\sim 2\times10^{11}$ cm$^{-2}$ have been reported in samples that were undoped,[37] meaning that PC-transferred graphene films approach intrinsic doping. A lack of polymer strands, as seen in scanning electron microscopy images (see SI), suggests that PC is cleaner than other polymer scaffolds.

Raman spectroscopy has proven to be a powerful, non-destructive tool for assessing the vibrational and electronic properties of carbon-based nanomaterials. For graphene layers, there are three major Raman bands called the D, G, and 2D (also known as G') bands, respectively. These bands' positions and full-width at half maximum (FWHM) values determine information about layer number, doping, and strain in the graphene films.[38-42] In Fig. 2, we give our Raman statistics for G and 2D band positions and FWHM for the PMMA, PMMA/PC, and PC transfer scaffolds. Table 1 also summarizes the statistics for several polymer scaffolds extracted by Gaussian fitting the Raman parameters. In the SI, we account for all of these factors and develop a general empirical model for our small-grain CVD samples.

The 2D band FWHM is quite sensitive to strain,[38, 40, 41] and in Fig. 2d we can decouple the strain (magenta, softened FWHM) and doping (cyan, stiffened FWHM) contributions. Following the model detailed in the SI, we find a strain in the PMMA films of $\varepsilon = -0.19 \pm 0.07\%$ and a doping increase of $\Delta n = (1.59 \pm 0.03) \times 10^{13}$ cm$^{-2}$. Instead of the expected p-type behavior for PMMA, we find an overall n-type doping in the films. This likely results from graphene encapsulation of water decorated silanol groups,[43, 44] as observed morphologically in the supplement. Furthermore, the doping observed in Raman spectroscopy is the total doping in films and therefore is subject to co-doping by p-type species like PMMA. Figures 2e-h show the Raman metrics for graphene transferred with a PMMA/PC bilayer. Following the previous discussion, we find a strain of $\varepsilon = -0.18 \pm 0.06\%$ and doping of $\Delta n = (1.40 \pm 0.03) \times 10^{13}$ cm$^{-2}$, again n-type. The PMMA/PC-transferred and the PMMA-transferred graphene differ in doping by $(1.90 \pm 0.44) \times 10^{12}$ cm$^{-2}$, a small difference in cleanliness. Nevertheless, we cannot conclude at that this difference is statistically significant at 99% confidence.

Finally, Figs. 2i-l give the PC-transferred graphene film's Raman metrics. The strain in the PC-transferred films (magenta, 2D) is $\varepsilon = -0.27 \pm 0.07\%$ with doping of $\Delta n = (2.00 \pm 0.04) \times 10^{13}$ cm$^{-2}$. Despite the heightened strain, the PC-transferred films have a higher n-type doping (i.e., lowered co-doping) than the PMMA- and PMMA/PC-transferred graphene films. This results in a doping difference of $(4.10 \pm 0.52) \times 10^{12}$ cm$^{-2}$ between the films. At a 99% confidence level, these two populations are statistically different, caused by the inherent p-type doping in the PMMA-transferred graphene films. Previous reports gave a p-type doping due to the PMMA of $p \sim 10^{12}$ cm$^{-2}$ (electrical)[12, 16] and ~$1.6 \times 10^{12}$ cm$^{-2}$ (for 495K molecular weight),[45] both consistent with the ~$4 \times 10^{12}$ cm$^{-2}$ change we are seeing here between PC- and PMMA-transferred graphene. Hence, these Raman data show the PMMA-induced p-doping occurring in transferred films.

Directly observing the polymer residues at relevant, micron-sized length scales is important for assessing how severe the contamination is for electronic devices, encapsulation layers, and other graphene applications. First, in Fig. 3a we show an atomic force microscopy (AFM) image for PMMA-transferred graphene on SiO$_2$ after a 2 hr Ar/H$_2$ forming gas anneal at 400°C. The surface of Fig. 4a is smooth post anneal, with a 0.33 nm root-mean-square (RMS) roughness in the blue box. Note that we give RMS roughness values for the subsequent images in the captions of Figs. 3 and 4, respectively. Despite the clean surface of Fig. 4a, the depolymerization of PMMA by thermal degradation and bond scission[46] is inherently inhomogeneous. To demonstrate this, we show an AFM image of a different area of the same annealed sample in Fig. 4b. The surface is quite rough from incompletely removed PMMA strands. This inhomogeneity, coupled with the deleterious chemical[15, 30] and electrical effects[47] that happen after Ar/$H_2$ annealing, illustrates the need for a room temperature polymer removal method.

Figure 3c demonstrates the level of surface contamination that occurs without annealing PMMA-transferred graphene in forming gas. PMMA strands thoroughly decorate the graphene/$SiO_2$/Si surface, and the tip's image convolves with larger PMMA debris. Comparatively, the PC-transferred graphene in Fig. 3d is remarkably cleaner without a thermal anneal. Both films in Figs. 3c and 3d come from the same graphene growth. Outside of growth-related morphological features,[9-11] Fig. 3d is featureless, suggesting that the room temperature dissolution of PC off graphene in chloroform is effective. Figures 3e-g show AFM images for PMMA-, PMMA/PC-, and PC-transferred graphene films on $SiO_2$, all from the same growth with the polymers dissolved in chloroform simultaneously. We give line profiles below the images, all taken along the inset gray dotted lines. The sample in Fig. 3e is less rough compared to Fig. 3c, as chloroform solvates polymers better than dichloromethane and acetone (see SI). Nevertheless, there is a still a sizable amount of PMMA residue on the sample, as demonstrated by a higher RMS roughness and the jagged line profile. Nevertheless, the PMMA/PC bilayer and PC scaffolds in Figs. 3f-g have lower RMS roughness values and smoother line profiles than the PMMA scaffold in Fig. 3e. This gives additional evidence that scaffolds with PC layers in contact to the graphene are sufficiently removed at room temperature.

We note that the wrinkle density in the PC-transferred film of Fig. 3g is high, caused by the thin (~70 nm, see SI) PC scaffold used. These wrinkles are mitigated with the stronger PMMA/PC bilayer in Fig. 3f. We also took AFM data of PLA- and PPA-transferred films on $SiO_2$ (see SI). In brief, both transfers leave residue on the graphene surface and are highly sensitive to their substrate interactions, consistent with the results of Figs. 1 and 2.

We also make use of our PC transfer process in the heterogeneous layering of graphene, fluorinated graphene (FG), and CVD hexagonal boron nitride (h-BN). In Figs. 4a and 4b, respectively, we show AFM images for one layer (Fig. 4a) and two layers (Fig. 4b) of PMMA-transferred graphene. The surface of Fig. 4a is akin in morphology to Fig. 3e. In Fig. 4b, when we wet-transfer[24] a second PMMA-based graphene layer, water is trapped at the graphene-graphene interface. We do not see intercalated chloroform in our samples (see SI), contrasting a recent study.[48] The water gives a rough morphology affected by the remnant hydrophobic PMMA strands, resulting in pinholes and no obvious water layering.[24] On the contrary, PC-transferred graphene results in a smoother morphology, as shown for one PC-transferred layer (Fig. 4c) and for two PC-transferred layers (Fig. 4d). Water is again trapped at the graphene-graphene interface of Fig. 4d, forming filaments and layers[24, 48] and not an amorphous film. Water layering is only possible if graphene's wetting properties are preserved, whereby the $SiO_2$/Si substrate templates the water through the graphene.[49] Hence, the PC-transferred graphene films leave insufficient residue to affect those wetting properties and disrupt water layering.

To examine these wetting phenomena in more detail, we layer different low-dimensional nanomaterials with PC. Figure 4e gives an AFM image a PC-transferred (no annealing) graphene/water/graphene stack. In Fig. 4f, we show an AFM section of a PC-transferred CVD graphene layer on top of a FG layer (for fluorination details, see SI).[10] Here, the superhydrophobic FG layer disrupts the graphene wetting transparency[49] and brings about point-like water accumulation. Without PC, these hydration characteristics would be obscured. h-BN is also hydrophobic, and we PC-transfer one layer (Fig. 4g) and two layers (Fig. 4h) of CVD h-BN. Like the graphene/$H_2O$/FG stack, the entrapped water is point-like, from the hydrophobic h-BN. h-BN transfer must take place with PC, as a forming gas Ar/$H_2$ "clean" attacks h-BN.[44]

Figure 4i summarizes the RMS roughness values from several AFM measurements of PMMA, PC, annealed PMMA, and photoresist-transferred graphene films. While the Ar/$H_2$ annealed PMMA-transferred films have the lowest RMS roughness ($\Delta = 2.9 \pm 0.4$ Å), PC-transferred films are also fairly smooth ($\Delta = 5.1 \pm 0.8$ Å). Nevertheless, the inhomogeneous PMMA removal, more pronounced graphene-substrate interaction,[47] and covalent PMMA re-hybridization[15] introduced by annealing make it less desirable. Moreover, graphene's temperature sensitive applications, like those involving biomolecule encapsulation or flexible substrates, prohibit annealing. Non-annealed PMMA transfers have a four-fold higher RMS roughness ($\Delta = 23.4 \pm 4.1$ Å) than non-annealed PC transfers, and AZ5214-transferred films are marginally smoother ($\Delta = 8.6 \pm 1.4$ Å) yet brittle (see SI).

Finally, we give atomic scale evidence of the cleanliness of PC-transferred graphene. In Figs. 4j-l, we show atomic resolution ultrahigh vacuum scanning tunneling microscopy (UHV-STM) images of PC-transferred graphene at two different degas[24, 25] temperatures. Figure 6j reveals a STM derivative image of non-annealed, PC-transferred graphene on 90 nm $SiO_2$/Si, degassed at ~54°C. Atomic structure of the graphene is present, albeit noisy. We note that the PMMA bond scission is not possible at this temperature.[46] As determined from the previous AFM images, equivalently prepared, PMMA-transferred samples have autocorrelation lengths of ~10 nm (25 μm²). Comparatively, PC-transferred samples have autocorrelation lengths greater than ~100 nm (25 μm²). The PMMA autocorrelation length is well within the AFM tip's radius of curvature and does not necessarily imply clean graphene regions between PMMA strands. Thus, the probability of serendipitously encountering an atomically clean graphene region for PMMA-transferred samples *via* STM is low. Figs. 4k and 4l show a STM topograph and derivative image, respectively, for PC-transferred graphene on mica, degassed at ~130°C. The scan shows improved resolution and graphene's atomic lattice is evident. The higher temperature degas likely removes adsorbed water, thereby improving the surface resolution. Regardless, PMMA bond scission[46] or sublimation[16] do not occur at ~130°C, further confirming the atomic-level cleanliness of the PC transfer.

We now comment on the mechanism of PC removal compared over and against PMMA, PLA, PPA, and other transfer scaffolds. An atomically clean graphene surface depends on the graphene-polymer interfacial adsorption and charge transfer, as well as the polymer's molecular weight and reactivity. For aromatic polymers like PC and PANI, the presence of electron withdrawing groups (EWGs) or electron donating groups (EDGs)[50] within the polymer's repeat unit will affect the interfacial graphene-polymer adsorption and charge transfer. Despite its nature as a conductive polymer, PANI-based scaffolds show strong doping in graphene (see SI). This occurs from the nitrogen moiety functioning as a strong EDG, donating its electron pair into the aromatic ring by resonance. Such charge transfer increases the polymer's adsorption energy on graphene at room temperature, like perylene-3,4,9,10-tetracarboxylic dianhydride (PTCDA).[51]

Conversely, the ester linkage within PC functions as a weak EDG in the repeat unit's aromatic rings via competing resonance and induction. Thus, this slight charge transfer in the aromatic rings of PC should drive adsorption by a "medium-range π–π*" electrostatic attraction with graphene, similar to benzoic acid on graphite.[52] The ordered interface allows for more effective polymer dissolution mechanically. PPA, which has an aliphatic backbone with pendant aromatic rings, is weaker in its EDG compared to PC. Compared against PANI and PC, PPA exhibits a smaller π–π electrostatic interaction with graphene, but it can share charge by its secondary ether linkage. Lacking an aromatic core, non-conjugated, aliphatic PLA and PMMA do not interface with graphene via $\pi$–$\pi$ interactions and theoretically more weakly adsorb. In turn, they should be easier to dissolve.

In addition to the graphene-polymer interfacial chemistry, the bulk polymer's molecular weight (MW) has a critical effect on its ultimate removal off graphene. To first order, high MW polymers have a larger area footprint on graphene, increasing the interaction probability with graphene features having high adsorption energy (i.e., grain boundaries).[25] Dissolving high MW polymers off graphene requires permeation through a gelling layer and disentanglement.[53] This process is inversely proportional to the MW,[53] provided there are no strong adsorption sites. Our high MW PMMA (~495K) is invariant to possible depolymerization without thermal bond scission, as PMMA is an aliphatic addition polymer.[15, 46] Recent reports[26, 51] also discovered that PMMA exposure to FeCl$_3$, a common Cu etchant, made the PMMA harder to remove, as Fe$^{+3}$ in acidic media potentially promotes PMMA cross-linking.[54] Our PMMA's high MW and acid exposure can help explain its difficult removal.[12, 15, 16, 30]

Furthermore, our aliphatic PLA films, despite being lower in MW (~55.4K), do not fully remove off graphene, even when heated past the gasification temperature.[32] This lack of depolymerizaton highlights increased graphene-polymer interaction. Conversely, as an aromatic, condensation polymer, PC can partially depolymerize via acid-induced hydrolysis, lowering the MW and promoting effective dissolution. This raises the question of how the acid would reach the PC films during transfer. Since PPA is an acid-sensitive polymer,[50] its depolymerization during the FeCl$_3$ Cu etching step can serve as an indicator of present, permeable acid vapor. In our supporting movies, we show rapid dissolution of a PPA/graphene sample, with PC/graphene and PMMA/graphene controls left intact. Thus, there is substantial acid vapor, enough to appear to lower the MW of PC from its starting 45K to 1-2K (see SI). PC's lowered MW, combined with the quasi-ordered, "medium-range $\pi$–$\pi$*" interaction for PC-graphene, makes its removal more effective than aliphatic PLA and PMMA.

We will note we transferred graphene with PMMA at 4K, a low MW similar to partially depolymerized PC oligomers (see SI). Interestingly, low MW PMMA comes off graphene cleanly, as observed by AFM (see SI). Since PMMA does not interact with graphene via $\pi$–$\pi$ interactions (i.e., no strong adsorption), the requisite condition for effective PMMA dissolution is a low enough MW to avoid entanglement and adsorption on graphene morphology. For medium to weakly interacting polymers, it appears that MW values less 10K are required for clean transfer. However, graphene supported by lower MW scaffolds suffers mechanical failure during transfer; the scaffolds must have a high MW overlayer (e.g. 495K PMMA). Conveniently, PC scaffolds partially depolymerize from the polymer's reactive character as a condensation polymer, obviating the need for an overlayer support, though one could be used.

In summary, we show room temperature, atomically clean graphene transfer onto $SiO_2$ and mica using PC as a transfer scaffold. We remove the PC scaffold at room temperature with chloroform, and no aggressive forming gas annealing is necessary to eliminate PC-based residues. PC transfers are significantly cleaner than the typical PMMA support and alternative scaffolds using PLA, PPA, PANI, TRT, and AZ5214 photoresist. We confirm the cleanliness of our PC transfer method against the alternative polymers by a thorough number of multi-scale characterization methods. PC-transferred films enable the heterogeneous layering of CVD graphene, FG, and CVD h-BN. Compared next to PMMA-transferred films, PC-transferred films preserve atomic interfaces and allow for the homogeneous layering of trapped water. We find that effective, room temperature removal of the scaffold off graphene requires a low molecular weight polymer with "medium-range" graphene-polymer interfacial interactions. PC fulfills all of these criteria, whereas the other polymers do not. PC-transferred films will also enable nanomaterial applications that are inherently more sensitive, such as graphene on flexible substrates, graphene as a biomaterial encapsulatory membrane, or CVD h-BN on arbitrary surfaces.

## Acknowledgment

This work has been partly sponsored by the U.S. Office of Naval Research (ONR) under grant N00014-13-1-0300 and the National Science Foundation (NSF) under grant CHE 10-38015. J.D.W. gratefully acknowledges funding from the National Defense Science and Engineering Graduate Fellowship (NDSEG) through the Army Research Office (ARO), the Beckman Foundation, and the Naval Research Enterprise Intern Program (NREIP). We are indebted to A. Liao, A. Rangarajan, S. Schmucker, and D. Estrada for helpful discussion and insights. We kindly acknowledge K. Chatterjee and R. Mehta for assistance during graphene transfer.

**Supporting Information**

The supplement contains additional discussion of the materials and methods, experimental data, and movies. This document is available online free of charge.

**Table 1.** Graphene Raman mapping statistics for different transfer scaffolds.

| Polymer | $\omega_D$ (cm$^{-1}$) | $\Gamma_D$ (cm$^{-1}$) | $\omega_G$ (cm$^{-1}$) | $\Gamma_G$ (cm$^{-1}$) | $\omega_{2D}$ (cm$^{-1}$) | $\Gamma_{2D}$ (cm$^{-1}$) | I(2D)/ I(G) | I(D)/ I(G) |
|---|---|---|---|---|---|---|---|---|
| PMMA | 1333.2±13.1 | 15.6±21.4 | 1599.9±2.5 | 9.1±3.3 | 2652.3±3.3 | 29.6±6.4 | 1.29±0.39 | 0.12±0.11 |
| PMMA/PC | 1333.2±5.3 | 23.4±17.0 | 1600.3±1.9 | 13.0±2.8 | 2657.9±3.3 | 32.5±2.5 | 1.96±0.32 | 0.07±0.03 |
| PC | 1329.1±4.9 | 16.6±14.6 | 1599.3±3.5 | 10.6±2.3 | 2653.9±4.2 | 31.9±1.8 | 1.60±0.18 | 0.11±0.04 |
| PLA | 1327.6±8.5 | 15.0±1.9 | 1592.5±3.2 | 18.1±2.1 | 2633.2±7.7 | 46.2±2.3 | 2.06±0.20 | 0.08±0.06 |
| PMMA/PANI | 1333.3±2.1 | 17.3±2.9 | 1603.6±1.7 | 18.6±4.7 | 2660.3±1.9 | 29.9±1.6 | 1.09±0.11 | 0.17±0.08 |

**Figures**

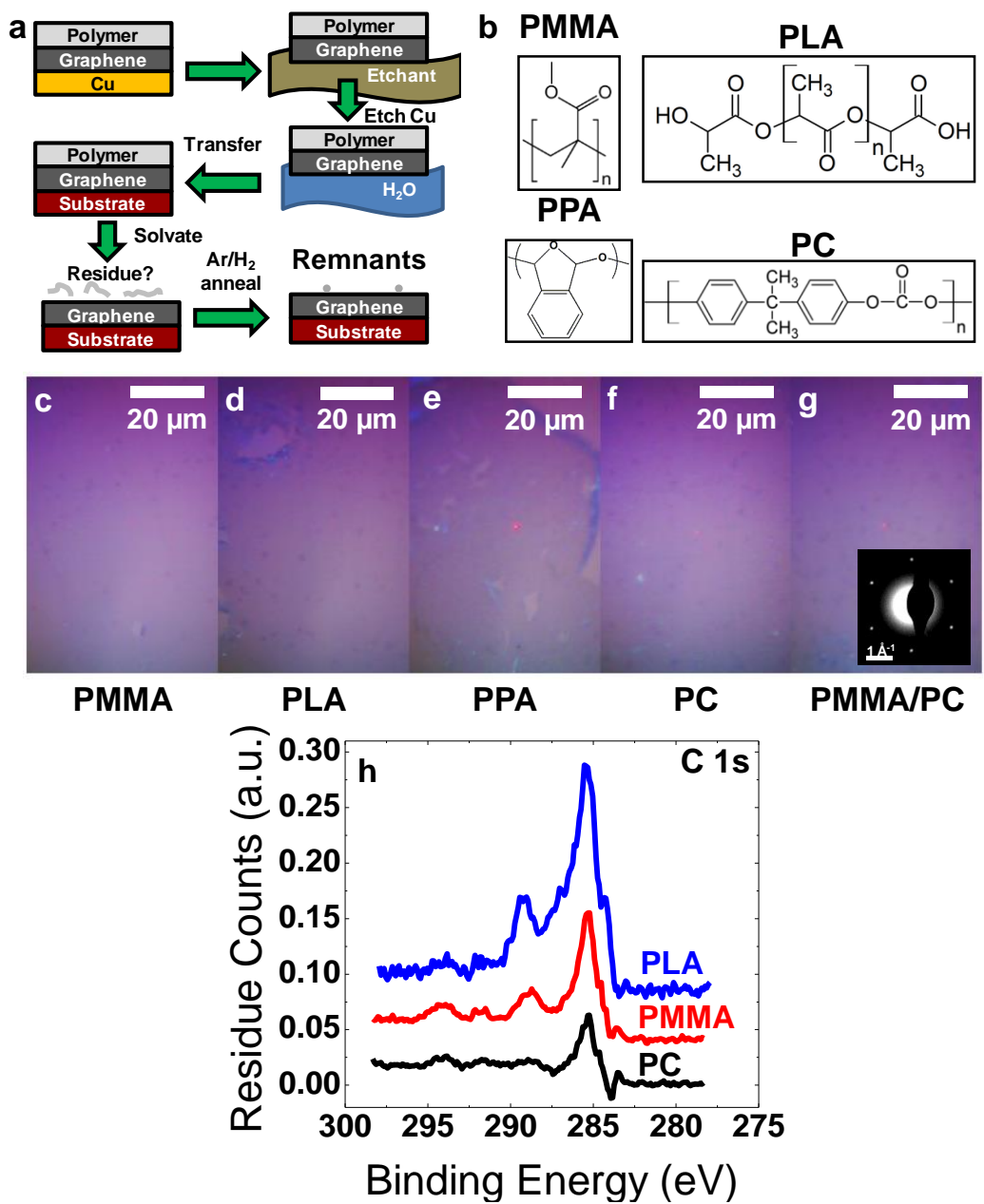

**Figure 1.** Polymer residues introduced by the chemical vapor deposition (CVD) graphene transfer process. **(a)** CVD graphene transfer process flow. Polymer residues often remain after a forming Ar/H$_2$ gas anneal. **(b)** Chemical formulae for the different polymers used. Optical images of large-area graphene transferred by **(c)** poly(methyl methacrylate) (PMMA), **(d)** poly(lactic acid) (PLA), **(e)** poly(phthaldehyde) (PPA), **(f)** poly(bisphenol A carbonate) (PC), and **(g)** PMMA/PC bilayer polymer scaffolds. Inset in (g) gives a representative, monolayer graphene diffraction pattern transferred by bilayer PMMA/PC. **(h)** C 1s photoelectron spectra after main graphene peak subtraction (sp$^2$ and sp$^3$) for PLA-, PMMA-, and PC-transferred graphene on SiO$_2$/Si. The residue counts are lowest for PC transfers.

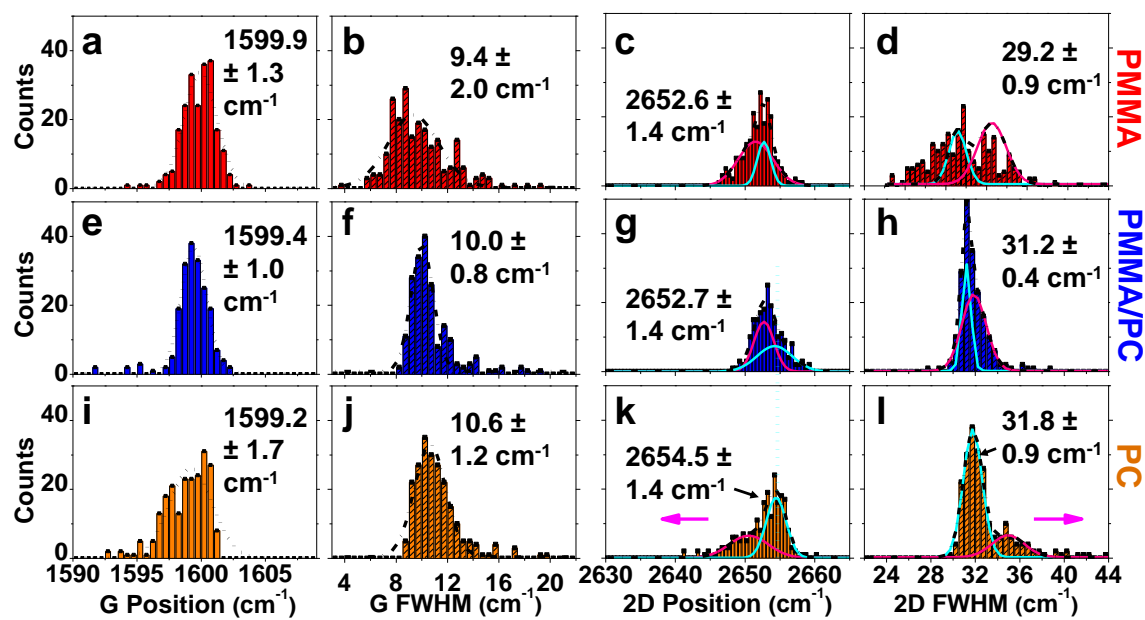

**Figure 2.** Raman spectra maps for graphene transferred by different polymer scaffolds. Distributions of G band position (**a**), G band full-width at half maximum (FWHM) (**b**), 2D band position (**c**), and 2D band FWHM for graphene transferred with PMMA. Mean and standard deviations listed, and strain-based contributions to the distribution are fit with magenta Gaussians. Distributions of G band position (**e**), G band FWHM (**f**), 2D band position (**g**), and 2D band FWHM (**h**) for graphene transferred with PMMA/PC (PC contacting graphene). Distributions of the G band position (**i**), G band FWHM (**j**), 2D band position (**k**), and 2D band FWHM (**l**) for graphene transferred with PC. Doping is lowest with PC, and the PC and PMMA populations are statistically different at 99% confidence level.

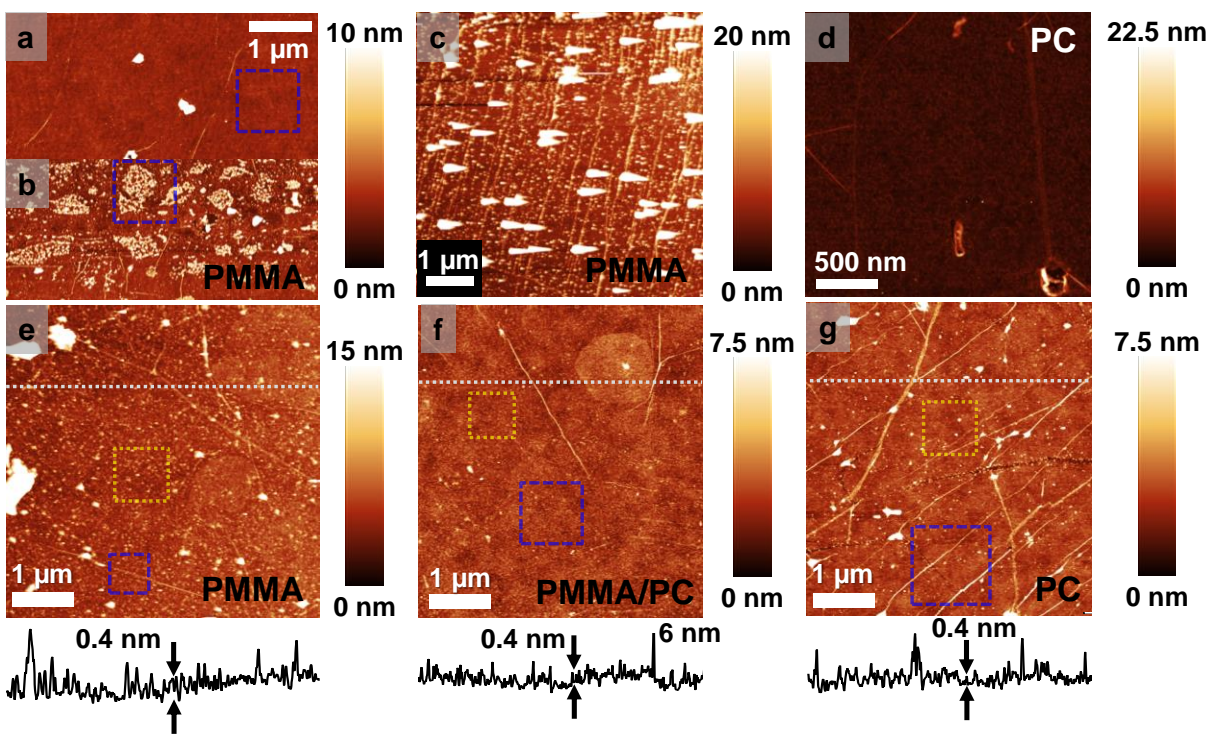

**Figure 3.** Changes in graphene surface morphology caused by the transfer polymers. AFM images **(a,b)** of PMMA-transferred graphene after a 90 min, 400°C Ar/H$_2$ anneal. While the region in (a) is clean, considerable, partially depolymerized PMMA remains on the graphene in (b). RMS roughness values: 0.33 nm (box, *a*), 1.72 nm (image, *a*), 2.8 nm (box, *b*), and 3.58 nm (image, *b*). AFM images of PMMA-transferred **(c)** and PC-transferred **(d)** graphene (same growth) without a forming gas anneal. The PC graphene surface is markedly smoother. Morphologies of PMMA-transferred **(e)**, PMMA/PC-transferred **(f)**, and PC-transferred **(g)** graphene films, all transferred from the same growth material. Both PMMA/PC and PC films are cleaner than PMMA films, with PMMA/PC films having fewer transfer-induced wrinkles. The overlaid grey lines correspond to the line profiles below each respective image. RMS roughness values: 1.21 nm (yellow, *e*), 1.05 nm (blue, *e*), 3.58 nm (image, *e*), 0.57 nm (yellow, *f*), 0.62 nm (blue, *f*), 0.71 nm (image, *f*), 0.71 nm (yellow, *g*), 1.07 nm (blue, *g*), and 1.76 nm (image, *g*).

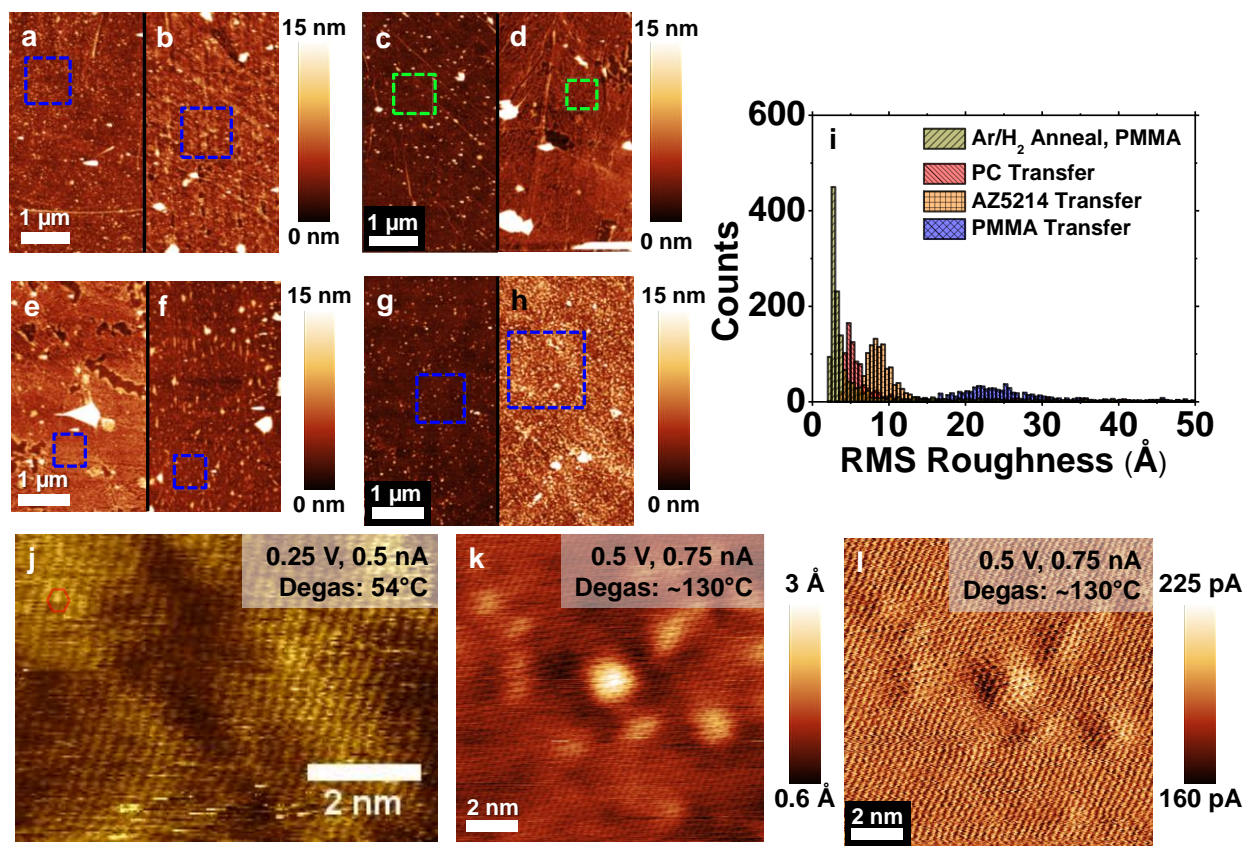

**Figure 4.** PC-enabled layering of low dimensional nanomaterials and atomic cleanliness of PC removal. One layer (**a**) and two layer (**b**) PMMA-transferred graphene on $SiO_2/Si$. The graphene layers are separated by trapped water, which does not layer from PMMA contamination. RMS roughness values: 1.4 nm (box, *a*), 1.87 nm (image, *a*), 1.65 nm (box, *b*), and 2.51 nm (image, *b*). One layer (**c**) and two layer (**d**) PC-transferred graphene on $SiO_2/Si$. In (d), water forms layers and tendrils between the graphene layers due to a PC-enabled, clean surface. RMS roughness values: 0.94 nm (box, *c*), 1.99 nm (image, *c*), 1.10 nm (box, *d*), and 11.5 nm (image, *d*). (**e**) Water trapped between two PC-transferred graphene (G) layers. (**f**) Water trapped between PC-transferred graphene on a fluorinated, PC-transferred graphene (FG) layer. There is water layering in the G/$H_2O$/G stack, compared to point-like, trapped water in G/$H_2O$/FG stack. RMS roughness values: 0.9 nm (box, *e*), 4.7 nm (image, *e*), 1.1 nm (box, *f*), and 5.4 nm (image, *f*). One layer (**g**) and two layers (**h**) of PC-transferred, CVD hexagonal boron nitride (h-BN). Like the G/$H_2O$/FG stack, point-like water accumulates between the layers in (h). RMS roughness values: 0.93 nm (box, *g*), 1.7 nm (image, *g*), 2.4 nm (box, *h*), and 3.2 nm (image, *h*). (**i**) RMS roughness histogram for different transfer scaffolds, revealing that PC-transferred graphene is as smooth as Ar/$H_2$ annealed graphene. (**j**) Atomic resolution scanning tunneling microscopy (STM) derivative image of a PC-transferred graphene film on $SiO_2/Si$, degassed at 54°C. Atomic resolution STM topograph image (**k**) and current image (**l**) for PC-transferred graphene on mica, degassed at ~130°C.

*Supporting Information*

# Annealing Free, Clean Graphene Transfer using Alternative Polymer Scaffolds

Joshua D. Wood[1,2,3], Gregory P. Doidge[1,2,3], Enrique A. Carrion[1,3], Justin C. Koepke[1,2], Joshua A. Kaitz[2,4], Isha Datye[1,2,3], Ashkan Behnam[1,3], Jayan Hewaparakrama[1,3], Basil Aruin[1,2], Yaofeng Chen[1,2], Hefei Dong[2,4], Richard T. Haasch[5], Joseph W. Lyding,[1,2*] and Eric Pop[6*]

*[1]Dept. of Electrical & Computer Eng., Univ. Illinois at Urbana-Champaign, Urbana, IL 61801*
*[2]Beckman Institute, Univ. Illinois at Urbana-Champaign, Urbana, IL 61801*
*[3]Micro and Nanotechnology Lab, Univ. Illinois at Urbana-Champaign, Urbana, IL 61801*
*[4]Dept. of Chemistry, Univ. Illinois at Urbana-Champaign, Urbana, IL 61801*
*[5]Materials Research Laboratory, Univ. Illinois at Urbana-Champaign, Urbana, IL 61801*
*[6]Dept. of Electrical Engineering, Stanford Univ., Stanford, CA 94305*

**Contents:**

[*] Correspondence should be addressed to lyding@illinois.edu and epop@stanford.edu

## Section S1. Materials and methods

*Chemical vapor deposition (CVD) of graphene on Cu*

To remove an anti-oxidation surface layer and reduce spurious nucleation sites,[1] we pre-cleaned our Alfa Aesar Cu foils in 10:1 $H_2O$:HCl for at least 3 min. We then rinsed excess HCl off the Cu and carefully dried the foils in $N_2$. We mounted the foils onto a cleaned quartz boat and annealed them at 1000°C in an Atomate CVD furnace using a 1:1 Ar:$H_2$ flow for 30 min. We only used Cu growth tubes that underwent a minimal number of growth runs, as backflowing Cu vapor from previous growths affected the growth quality. We then grew graphene for 25 min at 1000°C with 75 to 100 sccm of $CH_4$ and 50 sccm $H_2$. Samples were cooled under Ar, $CH_4$, and $H_2$, following previously established procedures.[2-4] These growths gave 90-95% monolayer coverage on the Cu surface with an approximate grain size of ~1 µm, as estimated by AFM, SEM, and Raman spectroscopy. Foils were stored under $N_2$ until used to mitigate Cu oxidation through graphene grain boundaries.[5]

*CVD of hexagonal boron nitride (h-BN) on Cu*

We also used 99.8% Alfa Aesar foils for the CVD of h-BN. The foils underwent the same pre-cleaning procedure. We grew h-BN by heated sublimation of ammonia borane ($NH_3$–$BH_3$, Sigma Aldrich) in a stainless steel ampule. The Cu substrate was annealed for 2 hrs under Ar/$H_2$ at 1000°C. After annealing, we grew h-BN in an Ar/$H_2$ background for 25 min, subliming the precursor at ~95°C. Additional details surrounding our CVD h-BN growth process will be published elsewhere.[6]

*Fluorination of graphene*

We fluorinated graphene with a Xactix silicon etcher at 1 T XeF$_2$ vapor pressure with a 35 T $N_2$ overpressure in normal (no pulse) mode. We fluorinated for 10 cycles at 60 s / cycle, consistent with previous work.[7, 8] These fluorination conditions are known to give highly fluorinated graphene (~$C_xF$, where $x < 4$).

*Poly(methyl methacrylate) (PMMA) transfer*

As detailed in the main text, we used a 495K A2 and 950K A4 PMMA bilayer (Anisole base solvent, 2% by wt. and 4% by wt., MicroChem) for PMMA-based transfer and overlayer support. Graphene on Cu (G/Cu) was cut to proper size and flattened by piranha cleaned glass slides. We coated each PMMA layer on G/Cu at 3000 RPM for 30 s, and then we cured each layer at 200°C for 2 min. More subtle details regarding the etching, cleaning, and ultimate graphene substrate transfer are given in the "Poly(bisphenol A carbonate) (PC) transfer" section below. After the PMMA/graphene film was on the substrate of choice, the PMMA was dissolved by chloroform solvation for at least 1 hr. Most polymer dissolution took place overnight, covered by a glass beaker. The samples were removed from the solvent and degreased with methanol, acetone, and IPA. Any optically obvious residues on the graphene chips were removed by further chloroform dissolution.

*Poly(bisphenol A carbonate) (PC) transfer*

We purchased poly(bisphenol A carbonate) (PC) from Sigma Aldrich (#181625, molecular weight of ~45K). We used the polymer as received. We dispersed PC in a chloroform (CF) solution at 1.5 wt. percent by volume, a more dilute weight percent than previous reports.[9, 10] We note that lower weight percents are imperative, because more concentrated PC solutions can gel during storage.[11] Amber-tinted bottles were used for solvent storage, as clear bottles lead to UV photodegradation of CF to phosgene. We also utilized dichloroethane as a solvent (see the samples in movie SM1) for PC, wherein we dissolved 3 wt. percent PC by volume in solution. DCE-based dispersions worked as well as CF-based ones, save the higher weight percent. For CF-based dispersions, very low weight percents (< 0.8 wt. percent) made the solutions less viscous, made the transfer scaffolds thinner (< 50 nm), and caused poor graphene transfer; therefore, we avoided solutions at lower weight percents. We added PC to a piranha cleaned amber bottle with chloroform and agitated the solution until no visible PC solid remained. Occasionally, we employed an additional 30 min sonication to more fully disperse the PC. The PC with chloroform solution was sealed with paraffin wax to avoid chloroform evaporative loss and concentration modification.

The G/Cu was placed on a spin coater with no additional support, and the PC films were spun onto the G/Cu at 3000 RPM for 30 s. We also spun PC at higher rates (5000 RPM and 7000 RPM for 30 s, each), giving a thinner polymer support. However, PC dissolution in solvent was not improved for these thinner PC films, and the structural support of these films was compromised. We performed no bake out of the solvent for PC samples, which is normally 200°C for 2 min for our PMMA-based transfers. For thicker PC scaffolds, we repeated the 3000 RPM for 30 s spin-coating process three times more (see Table S2).

We removed the backside graphene on the Cu by 90 W $O_2$ plasma operated at a throttle pressure of 100 mTorr for ~30 s. We optically assessed the top side of the film to ensure that the plasma did not degrade the polymer scaffold. We etched the Cu substrate overnight in a $FeCl_3$ etchant (Transene Co., CE-100), covered at room temperature. Occasionally, we etched the Cu with ammonium persulfate (Transene Co., APS-100), which etched more cleanly but produced bubbles on the PC/graphene film underside. These bubbles prevented further Cu etching and gave circular depressions within the transferred film (see Fig. S3).

After overnight etching, we raised the $FeCl_3$ etchant fluid level by careful DI water dilution. The raised fluid level made PC/graphene removal from the solution easier. With piranha cleaned glass slides, we wicked the graphene (see supporting information) out of the solution and onto the slides. We then cleaned residual etchant off the PC/graphene films by placing them on DI water for ~15 min. After this bath, we transferred (with glass slides) the PC/graphene films into a series of modified RCA cleaning baths.[12] The modified RCA cleaning baths were made up of SC-2 and SC-1, respectively, with the SC-2 composed of 20:1:1 $H_2O$:$H_2O_2$:HCl and the SC-1 composed of 20:1:1 $H_2O$:$H_2O_2$:$NH_4OH$. We cleaned PC/graphene in SC-2 for ~15 min. In early experiments, we transferred the films into a SC-1 bath for ~15 min. In later experiments, we determined that the $NH_4OH$ from the SC-1 gave adsorbed nitrogen on the underside of the graphene films. After identifying this, we eliminated the SC-1 cleaning step. We manipulated the PC/graphene films into a final DI water bath. From this bath, we transferred the films to the substrates of interest, usually piranha cleaned 90 nm $SiO_2$/Si. We also performed the RCA clean and an $O_2$ plasma descum (90 W for 15 min) on the $SiO_2$/Si substrates. Compared with the piranha clean, these procedures did not substantially affect the graphene transfer.

After the PC/graphene films were on the substrate, we spun off excess water from the graphene-substrate interface at 7000 RPM for 60 s. The competing capillary and centripetal forces prevented the PC/graphene films from delaminating from the substrate. We found that the spinning step was imperative for more hydrophobic substrates like H-passivated Si(100) or graphene already on $SiO_2$/Si. We then drove off additional water by placing the samples on a hot plate at 60°C for ~5 min. After that heating step, we ramped the hot plate[12] to 150°C and, when the 150°C temperature was reached, we held the samples there for ~5 min. We dissolved the PC scaffold in chloroform overnight. We then degreased the chips with methanol, acetone, and IPA, and we dried the chips with house $N_2$.

We note that our samples were incidentally exposed to UV light by ambient exposure before polymer dissolution. Thus, it is possible that UV-catalyzed chloroform, in a phosgene derivative, could proceed through transesterification with PC to produce additional phosgene and bisphenol A (BPA). While in the main text we argue for partial PC depolymerization by acid-based hydrolysis, the UV exposure during polymer liftoff could also partially depolymerize the PC. This would assist in the polymer removal, as it results in a lower molecular weight.

*PMMA/PC bilayer transfer*

PC layers were spun onto G/Cu first and not cured at elevated temperature. Approximately ~1 min after spinning, the PMMA bilayer was spun onto the PC/G/Cu structure. Curing and transfer then proceeded per the practice outlined above.

*Poly(lactic acid) (PLA) transfer*

For the sample shown in Figs. S7a-c, we transferred the graphene using ~1 g of 55.4K MW poly(lactic acid)[13] dissolved in ~25 mL of chloroform, giving a solution with a 2.7 wt. percent. For the sample in Fig. S7d, this solution was diluted 3:1 in chloroform. G/Cu samples were placed on a spin coater with no additional support, and the PLA films were spun onto the G/Cu at 3000 RPM for 30 s, regardless of the dilution. No sample bake out was performed, and the subsequent steps followed those detailed in the "Poly(bisphenol A carbonate)" section.

*Poly(phthalaldehyde) (PPA) transfer*

We purified O-phthalaldehyde (OPA, purchased from Alfa-Aesar) according to a literature procedure,[14] and we dried the sample under high vacuum for 24 hours. OPA (1.00 g, 7.5 mmol) is weighed into a Schlenck flask and dissolved in anhydrous dichloromethane (10 mL). The solution is cooled to $-78°C$ and boron trifluoride etherate is added (purchased from Sigma-Aldrich, 8 μL, 60 μmol). The reaction is left stirring at $-78°C$ for 2 hours, then acetic anhydride (purchased from Fisher, 0.25 mL, 2.6 mmol) and pyridine (purchased from Alfa-Aesar, 0.22 mL, 2.7 mmol) are added. The mixture is left stirring 2 hours at $-78°C$, then the polymer precipitated by pouring into methanol (100 mL). The product is collected by filtration, then re-precipitated from dichloromethane and washed in methanol and diethyl ether (0.84 g, 84%). $^1H$ NMR (500 MHz, DMSO-$d_6$) δ 7.85-7.00 ppm (br, 4H, aromatic), 7.00-6.20 ppm (br, 2H, acetal). $^{13}C\{^1H\}$ NMR (500 MHz, CDCl$_3$) δ 138.8 ppm, 130.2 ppm, 123.5 ppm, 105.0-101.8 ppm. We subsequently refer to this product as PPA.[15-18] GPC on PPA revealed a molecular weight (MW) of 27.1 kDa (polydispersity PDI = 2.09).

We mixed 0.16 g and 0.24 g of 27.1K PPA in ~15 mL and ~18 mL of chloroform, respectively. We refer to the 0.16 g solution as M1 and the 0.24 g solution as M2. The mixture was allowed to sit overnight, and it was sealed with wax. Three samples were made with PPA overlayers only (SA1, SB1, SC1). Samples SA1 and SB1 had M1 solution spun onto G/Cu substrates at 3000 RPM for 30 s. SA1 was etched using ammonical Cu (Transene Co., BTP) etchant, and SA1 contained undesirable intercalated contaminants from the etch. Sample SC1 had M2 solution spun onto G/Cu at 6000 RPM for 30 s. SB1 and SC1 were etched in ammonium persulfate (Transene Co., APS-100), and both samples structurally decomposed in one of the supporting movies (SM2).

All other PPA samples had M1 solution spun onto G/Cu at 3000 RPM for 30 s. These samples were coated with PMMA (495K and 950K) following our aforementioned procedure.

*Poly(aniline) (PANI) transfer*

We purchased emeraldine salt poly(aniline) (PC) from Sigma Aldrich (#556386, molecular weight of ~50K). We dissolved 0.193 g of poly(aniline) (PANI) in 10 mL of chloroform. This solution was agitated until the PANI was fully dissolved. All PANI samples were spun with this solution at 3000 RPM for 30 s with no solvent bakeout. PANI samples with a PMMA overlayer support had the PANI layer spun first, followed by a 495K and 950K PMMA layer. Solvent bakeout at 200°C was performed on the PMMA layers.

*Annealing*

We performed anneals on PMMA-based CVD graphene chips in 400 sccm Ar with 400 sccm $H_2$ for 90 min at 400°C. Both gases were of ultra-high purity (99.999% pure or better), minimizing graphene-based etching from gas contaminants.[19] If deemed necessary, we annealed PC-based CVD graphene chips in Ar/$H_2$ for 90 min at 450°C. To examine how water leaves the CVD graphene-CVD graphene and the CVD graphene-$SiO_2$/Si interface (Figs. S10 and S11), we annealed PMMA- and PC-based chips in Ar only for 60 min at 200°C. All anneals took place in an Atomate CVD furnace at atmospheric pressure with a throttled roughing pump configuration.

*Scanning Electron Microscopy (SEM)*

We used a FEI environmental SEM at 5 kV on graphene. All images were taken using a ultra high-definition mode, which increases the dwell time and the beam current. We maintained similar values for the brightness and contrast during image collection, so that the images in Fig. 2 can be adequately compared.

*X-ray Photoelectron Spectroscopy (XPS)*

A Kratos ULTRA XPS with a monochromatic Kα-Al X-ray line was used to collect data. We fitted all sub-peaks with Shirley backgrounds and Gaussian-Lorentzian (GL) mixing. The amount of GL character was optimized (*i.e.*, not fixed) in our fits, so as to lower the chi-squared value and be representative of the true chemical state of the sub-peak in question. All full-width at half maximum (FWHM) values were less than 3 eV. Charging effects on the sub-peak binding energy were corrected by offsetting to the Si 2p peak for $SiO_2$/Si. For the C 1s photoelectron, we employed the asymmetric Doniach-Sunjic (D-S) lineshape for the $sp^2$ carbon sub-peak.[20] All other sub-peaks were fitted using the aforementioned GL mixing procedure.

*Raman Spectroscopy*

We took most Raman spectra using a Renishaw Raman spectrometer at 633 nm excitation (~1-10 mW, ~2 μm spot) and inVia software. The acquisition time was 30 s, and the grating was 1800 lines/mm. During mapping, a 50X objective (~0.7 NA) was used, and the pixel-to-pixel distance was much larger than the spot size (~5 μm). To correctly identify the position of the D, G, and 2D bands from the mapping data, a Lorentzian fitting procedure was used, as detailed elsewhere.[4] For the graphene point Raman spectra shown in this document, we subtracted a polynomial background from the data, thereby lowering fluorescence. We then fitted the resultant data with Lorentzians using a Levenburg-Marquardt fitting procedure in Fityk. Occasionally, a Horiba Raman spectrometer was used (specifically, the PPA data in this document). The laser line was again 633 nm, and the power was kept below 10 mW. The acquisition time was again 30 s, and we used a 300 lines/mm grating and a 100X (~0.9 NA) objective in a backscattering geometry.

*Atomic Force Microscopy (AFM)*

We performed atomic force microscopy (AFM) measurements in tapping mode with ~300 kHz Si cantilevers on a Bruker AFM with a Dimension IV controller. Scan rates were slower than 2 Hz, and sampling is at least 512 samples per line by 512 lines. Most images were sampled at 1024 samples per line by 1024 lines. Images without substantial noise and stable phase imaging were selected for analysis. Images were de-streaked, plane fit, and analyzed using Gwyddion. RMS roughness values were determined by Gwyddion and through an algorithm written in MATLAB. Autocorrelation values were also determined and fit in Gwyddion. The AFM images shown in this document for graphene transferred with 4K molecular weight PMMA were taken on an Asylum Research MFP-3D AFM. On that system, tapping mode AFM was performed using ~300 kHz resonant frequency Si cantilevers (NSG30 AFM tips from NT-MDT).

*Device Transport*

Graphene was transferred onto 90 nm $SiO_2$/Si as previously described, using PMMA and PC based scaffolds. No annealing was performed. Source/drain electrodes (Ti/Au) and graphene channels were defined using a PMGI/PR stack and UV lithography. PMGI (MicroChem) was spun at 3500 RPM for 30 s and cured at 165°C for 5 min. Shipley 1813 PR (MicroChem) was spun on top of the cured PMGI at 5000 RPM for 30 s. The PR was soft baked at 110°C for 70 s, exposed to UV for 4 s on a Karl-Suss aligner (i-line) and developed for 50 s in MF-319 (MicroChem). In the case of electrodes, Ti (0.7 nm) and Au (40 nm) were e-beam evaporated followed by lift-off in hot n-methyl pyrrolidone (Remover PG, MicroChem). Channels were defined using an $O_2$ plasma RIE. Channel length (*L*) and width (*W*) ranged from 2 to 3 μm and 5 to 10 μm respectively. All measurements were performed in vacuum at room temperature with a Keithley 4200 Semiconductor Characterization System (SCS).

*Scanning Tunneling Microscopy (STM)*

Our experiments employed a homebuilt, room-temperature ultrahigh vacuum scanning tunneling microscope (UHV-STM) with a base pressure of $\sim 3 \times 10^{-11}$ Torr[21] and electrochemically etched W and PtIr tips.[22] We scanned the samples in constant-current mode to get topographic data. In this procedure, the tip height was feedback-controlled, maintaining a current set point while rastering the tip across the surface. We grounded the STM tip through a current amplifier, and we applied the tunneling bias to the sample. For the mica substrate in Figs. 6(k-l), we mounted a Si backing through which we could resistively heat the sample. Regardless, sample degasses occurred using a hot filament to heat the samples to ~54°C (thermocouple readout) and ~130°C.

*Gel Permeation Chromatography (GPC)*

Analytical gel permeation chromatography (GPC) analyses were performed on a system composed of a Waters 515 HPLC pump, a Thermoseparations Trace series AS100 autosampler, a series of three Waters HR Styragel columns (7.8' 300 mm, HR3, HR4, and HR5), and a Viscotek TDA Model 300 triple detector array, in HPLC grade THF (flow rate = 0.9 mL/min) at 25°C. The GPC was calibrated using a series of monodisperse polystyrene standards.

## Section S2. Transfer with thermal release tape (TRT) and AZ5214 photoresist

TRT-based transfers were previously reported for epitaxial graphene on C-face SiC[23] and for graphene on Cu.[24, 25] In these reports, the TRT-transferred graphene films often had transfer-induced holes in them from adhesion issues with the TRT, the graphene, and the substrate. We also see holes in our TRT transfers (Fig. S3), corroborating the adhesion concern. Some of these holes can be mitigated by hot press transferring.[25] Regardless, we observe significant sample-to-sample variability in the TRT transfers, resulting from inhomogeneities in the Cu growth substrate and from the TRT losing adhesive strength. Moreover, the TRT introduces contamination on the top-side of the graphene. Proper solvent treatment[23] can lower this doping but not eliminate it entirely. The adhesion and contamination issues make the TRT-based transfers less appealing.

AZ5214-based transfers are equally as holey as TRT transfers, and the scaffolds are more susceptible to mechanical breakage during transfer. In this transfer process, we coat and develop the AZ5214 PR onto the graphene on Cu following the procedures given in this document. Despite the PR development, we observe substantial contamination on the graphene caused by the PR (Fig. S4). This contamination, combined with the scaffold's poor structural integrity, make the AZ5214-based transfers intractable.

## Section S3. Strain and doping model for Raman spectra populations

In graphene-based Raman spectroscopy, the energy-dispersive D band originates from defects,[26] grain boundaries,[2, 3, 27] and edges[26, 28] within the graphene, and this band is centered about ~1335 cm$^{-1}$ ($E_L$ = 1.96 eV).[26, 29-31] Doubly-degenerate, $\Gamma$ point iTO phonons give rise to the G band at 1588 cm$^{-1}$ for intrinsic, monolayer graphene.[32] The energy-dispersive, two iTO phonon 2D band at 2645 cm$^{-1}$ (CVD, $E_L$ = 1.96 eV)[2] comes from a double resonance process between the $K$ and $K'$ valleys.[26, 29-31] Both the G and 2D bands are strongly affected by charge-transfer doping[32] and strain[33] in the graphene.

From previously published in-plane strain data on CVD graphene films,[34] we determine an expression for the 2D FWHM for compressive strains ($\varepsilon < 0$): $\Gamma_{2D}(\varepsilon) = (26.1 \pm 0.3) + (-33.2 \pm 1.4) \cdot \varepsilon$. This expression accounts for the standard error in the fit. Using this expression with the data in Fig. 2(d), we find a strain in the PMMA films of $\varepsilon = -0.19 \pm 0.07\%$. With this strain, we can estimate the graphene band shifts using proper Grüneisen parameters,[34] where the G band and 2D band shifts are 41.1 cm⁻¹/% and –72.3 cm⁻¹/%, respectively. Starting from ~1588 cm⁻¹ for the G band (typical of graphene on SiO₂/Si)[32] and ~2645 cm⁻¹ for the 2D band (at excitation $E_L = 1.96$ eV), our strain-shifted band positions are $\omega_G = 1580.0 \pm 2.8$ cm⁻¹ and $\omega_{2D} = 2659.0 \pm 4.8$ cm⁻¹, respectively. To arrive at the PMMA G band position at 1599.9 cm⁻¹ (Fig. 2(a) in the main manuscript), we must upshift the G band by $\Delta\omega_G = 19.9 \pm 3.0$ cm⁻¹.

We also can determine an empirical model that accounts for the strain-based increase[34] in the G band FWHM: $\Delta\Gamma_G(\varepsilon) = (-12.7 \pm 1.0) \cdot \varepsilon$. For the PMMA-based films, $\Delta\Gamma_G = 2.5 \pm 0.9$ cm⁻¹. The doping contribution appears high ($|n| > 5 \times 10^{12}$ cm⁻²) in all of the samples in Fig. 3, prohibiting G band electron-hole pairs for $E_F > \hbar\omega_G/2$ and making the doping contribution negligible.[35] Therefore, the G band FWHM reduces to non-electronic and strain-based contributions. Ubiquitous in our Raman spectra is an inhomogeneous G band broadening of approximately ~8 cm⁻¹, as previously noted.[32, 35] Combining the broadening with the strain increase, we arrive at a G band FWHM for the PMMA-transferred graphene films of $\Gamma_G = 10.5 \pm 0.9$ cm⁻¹, close to our measured value of $9.4 \pm 2.0$ cm⁻¹. This bolsters our proposed descriptions thus far regarding strain and doping in the PMMA-transferred film.

To reconcile the G band's position, we assign the aforementioned $19.9 \pm 3.0$ cm⁻¹ upshift required to doping in the PMMA-transferred graphene film. The upshift corresponds to a doping increase of $\Delta n = (1.59 \pm 0.03) \times 10^{13}$ cm⁻².[26] Herein, we have assigned the carrier type as n-type, for reasons that momentarily become evident. Analyzing the 2D band position allows us to assign the carrier type. Using the strain-shifted 2D band position of $2659.0 \pm 4.8$ cm⁻¹, we must downshift the band by $6.4 \pm 5.0$ cm⁻¹. The presence of a downshift implies n-type doping in the graphene.[29] Moreover, the approximate two-fold doping shift increase for the G band relative to the 2D band agrees well with the discrepancy in electron-phonon coupling for iTO phonons at the $\Gamma$ and $K$ points.[36] Thus, it appears that our room-temperature and 200°C annealed graphene films on 90 nm $SiO_2$/Si have trapped water under them, despite being rough.[2,37] Later in this document, we show that this trapped water n-type dopes the graphene from the electrostatic interaction between the Si–OH groups and the encapsulated water (*vide infra*).

We apply our model to the PMMA/PC bilayer of Figs. 2(e-h). From the model, we ascertain a compressive strain of strain of $\varepsilon = -0.18 \pm 0.06\%$, along with doping shifts of $\Delta\omega_G = 18.6 \pm 2.7$ cm$^{-1}$ and $\Delta\omega_{2D} = -5.0 \pm 4.6$ cm$^{-1}$, respectively. The G band upshift gives a doping in the graphene film of $\Delta n = (1.40 \pm 0.03) \times 10^{13}$ cm$^{-2}$, again n-type due to the entrapped water. Nonetheless, the lower doping concentration could result from p-type co-doping, resulting from co-mixed PMMA[20] in the PC interfacial layer. The PMMA/PC bilayer transferred graphene and the PMMA-transferred graphene of Figs. 2(a-d) differ in doping by $(1.90 \pm 0.44) \times 10^{12}$ cm$^{-2}$, as discussed in the main text. Indeed, when we test the difference between the two G band datasets (*i.e.*, Figs. 2(a) and 2(e), respectively), we cannot conclude at that they come from different populations at 99% statistical significance. Even though the PC contacts the graphene, it is possible that the PMMA partially co-mixes[20] during the 200 °C bakeout (*vide supra*).

Finally, we calculate the strain in doping present in our PC-based Raman data (Figs. 2(i-l)) using the aforementioned model. We find a strain of $\varepsilon = -0.27 \pm 0.07\%$ with doping of $\Delta n = (2.00 \pm 0.04) \times 10^{13}$ cm$^{-2}$ occurring from G and 2D band shifts of $\Delta\omega_G = 22.3 \pm 3.4$ cm$^{-1}$ and $\Delta\omega_{2D} = -10.0 \pm 5.3$ cm$^{-1}$, respectively. Water doping is again present, and the higher n-type behavior seen results from a lack of co-doping due to a cleaner graphene surface. Further comparisons between the PC, PMMA, and PMMA/PC films are made in the main text.

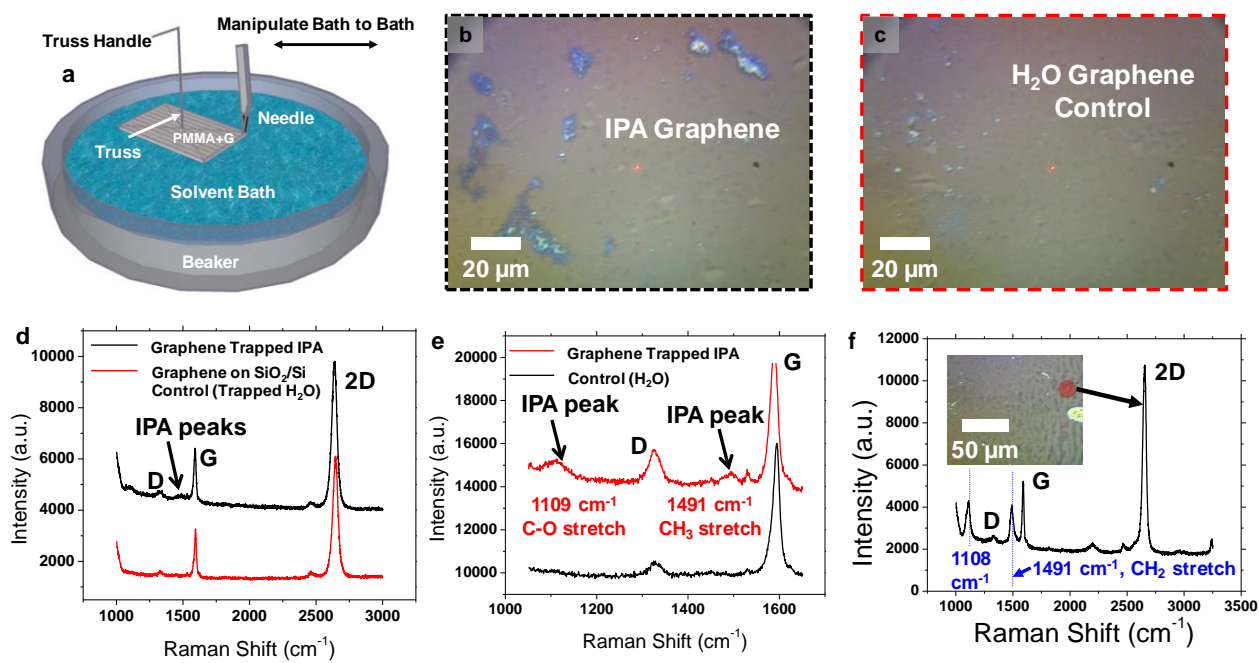

**Figure S1. PMMA-transferred graphene with different final solvent baths.** **(a)** Schematic of the PMMA scaffold transfer process used for transferring graphene into other final solvent solutions. Note that the high surface tension of water supports the graphene film during conventional wet transfer; the graphene does not float on the water. When the solvent is changed to something besides water, the lower surface tension causes the PMMA-graphene film to sink in the solvent. To counteract this, a metal truss is employed to prevent the PMMA-graphene from tumbling into solution. Optical images of graphene truss-transferred in 2-propanol **(b)** and in a $H_2O$ **(c)** control. **(d)** Point Raman spectra ($\lambda_{exc}$ = 633 nm, ~2 mW power, 50X, 30 s acquisition) for the optical images in (b) and (c), showing additional peaks from the IPA. **(e)** Zoom-in Raman spectra for the samples in (d). Peaks at 1109 $cm^{-1}$ and 1491 $cm^{-1}$ correspond to entrapped IPA solvent. The 1109 $cm^{-1}$ peak corresponds to the C–O stretching mode in IPA, and the 1491 $cm^{-1}$ peak corresponds to the $CH_3$ stretching mode.[36, 38, 39] **(f)** Point Raman spectrum for graphene truss-transferred in ethylene glycol. Optical image is shown inset. Graphene Raman peaks (D, G, and 2D) are present, and peaks at 1108 $cm^{-1}$ and 1491 $cm^{-1}$, respectively, correspond to C–O and $CH_2$ stretching modes.[40]

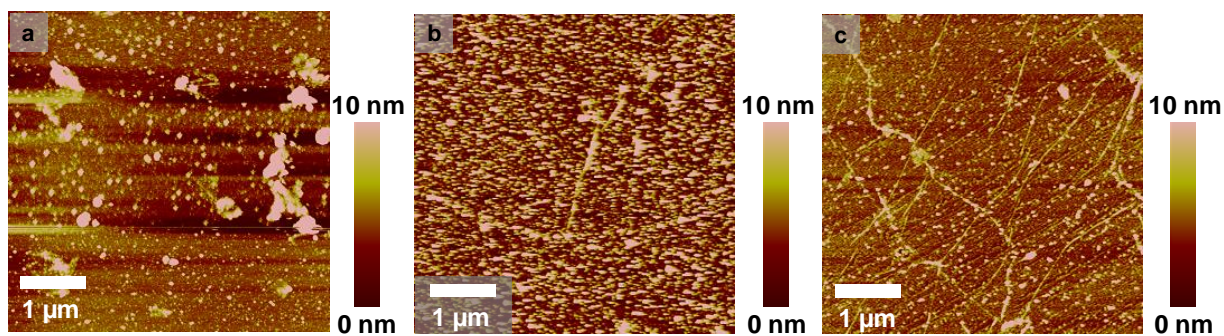

**Figure S2. Room-temperature removal of PMMA with different solvents and substrates**. Graphene grown at 1000°C for 25 min with 75 sccm $CH_4$ and 50 sccm $H_2$ on Cu using a pocketed approach.[7, 41] **(a)** AFM image of PMMA-transferred graphene on mica, with the PMMA partially removed by a 20 min acetone soak. **(b)** AFM image of PMMA-transferred graphene on $SiO_2$/Si, with the PMMA removed by a 20 min acetone soak. Compared to (a), the PMMA removal on $SiO_2$/Si is considerably lower. **(c)** AFM image

of PMMA-transferred graphene on mica, with the PMMA removed by a 20 min dichloromethane : methanol (1:1 ratio) soak. Dichloromethane appears effective in PMMA disentanglement.

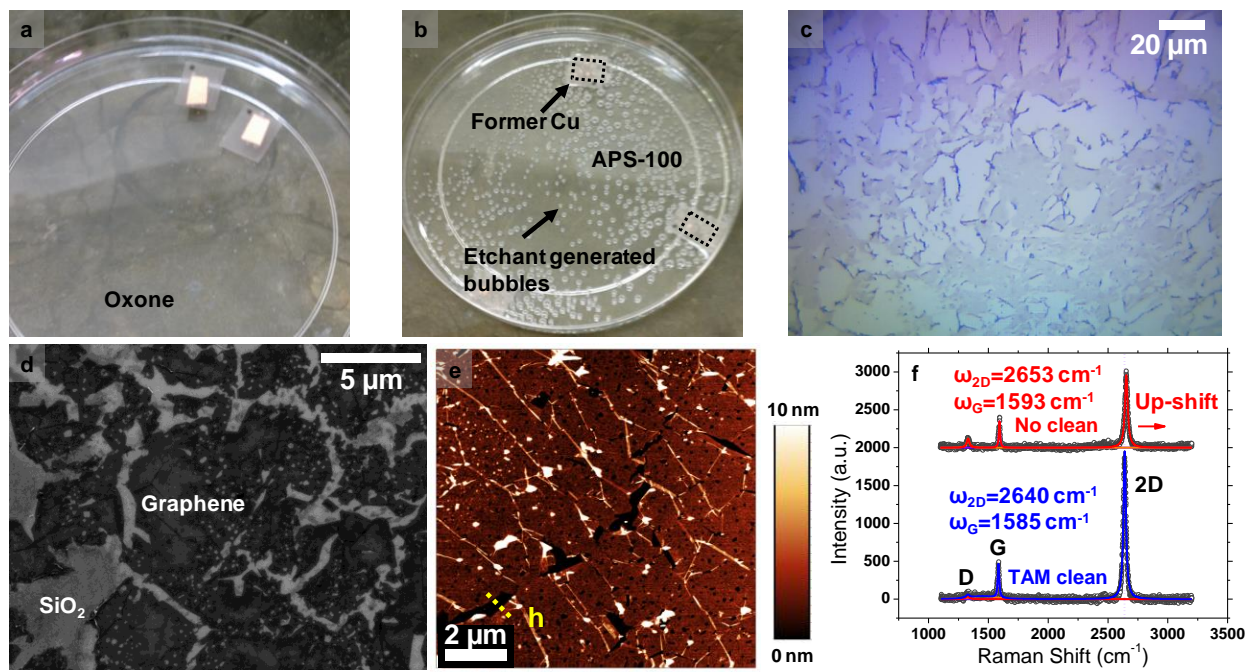

**Figure S3. Graphene transfer using thermal release tape (TRT). (a)** Photograph of Cu on thermal tape floating on potassium peroxymonosulfate (Oxone), a similar etchant to the commonly used ammonium persulfate. **(b)** Photograph of etched Cu in ammonium persulfate from Transene (APS-100). Etchant produces considerable amounts of bubbles due to the Cu reduction reaction. **(c)** Optical image of TRT-transferred graphene on $SiO_2$/Si. Many tears and scrolled edges are present. TRT-transferred film was released at 185°C and cleaned with a 10 min toluene : acetone : methanol bath. **(d)** SEM image of TRT-transferred graphene on $SiO_2$/Si. TRT with graphene is dried with $N_2$ before transferring to the $SiO_2$/Si. Poor tape release [25] and bad adhesion between the $SiO_2$/Si produces holes and tears within the transferred film. **(e)** AFM topographic image of TRT-transferred graphene on $SiO_2$/Si. Height from a tear edge to the substrate is $h = 3.0 \pm 0.6$ nm. Large height likely results from the backside graphene on Cu not being removed. Note that the circular, graphene-free depressions occur from bubbles produced by the etchant during transfer. **(f)** Point Raman spectra ($\lambda_{exc}$ = 633 nm, ~2 mW power, 50X, 30 s acquisition) for TRT-transferred graphene to $SiO_2$/Si for a sample with its TRT residues cleaned by a toluene : acetone : methanol clean (TAM clean) [8, 23] and for an unclean sample. The cleaned sample has considerably lower doping, as determined by the concurrent down shifts of the 2D and G bands.

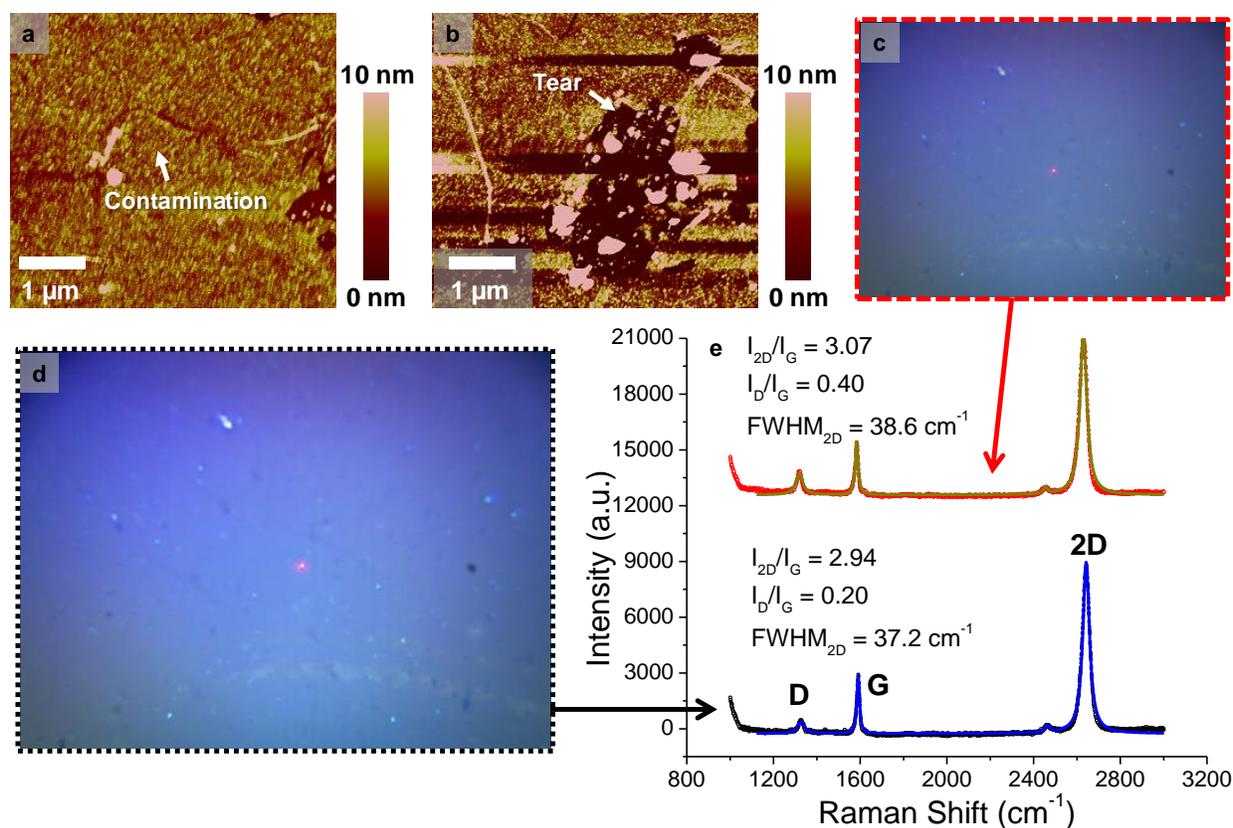

**Figure S4. Graphene transfer using AZ5214 photoresist**. AFM images of a continuous (**a**) and torn (**b**) graphene region on SiO$_2$/Si transferred by flood-exposed AZ5214 photoresist. 1 min UV flood exposure time used after the AZ5214 film was transferred to a SiO$_2$/Si chip, and then the film was developed in MF-319 developer. Both (a) and (b) show considerable contamination introduced by the photoresist, despite the flood exposure and development. Optical images (**c, d**) of the AZ5214 transferred film on SiO$_2$/Si. Tears apparent in both (c) and (d). (**e**) Point Raman spectra ($\lambda_{exc}$ = 633 nm, ~2 mW power, 50X, 30 s acquisition) corresponding to the optical images in (c) and (d). Both spectra are of monolayer graphene, but the large 2D FWHM (greater than 30 cm$^{-1}$) indicates strain in the film. The G FWHM values (not listed) are less than 15 cm$^{-1}$, revealing AZ5214 induced doping in the graphene film.

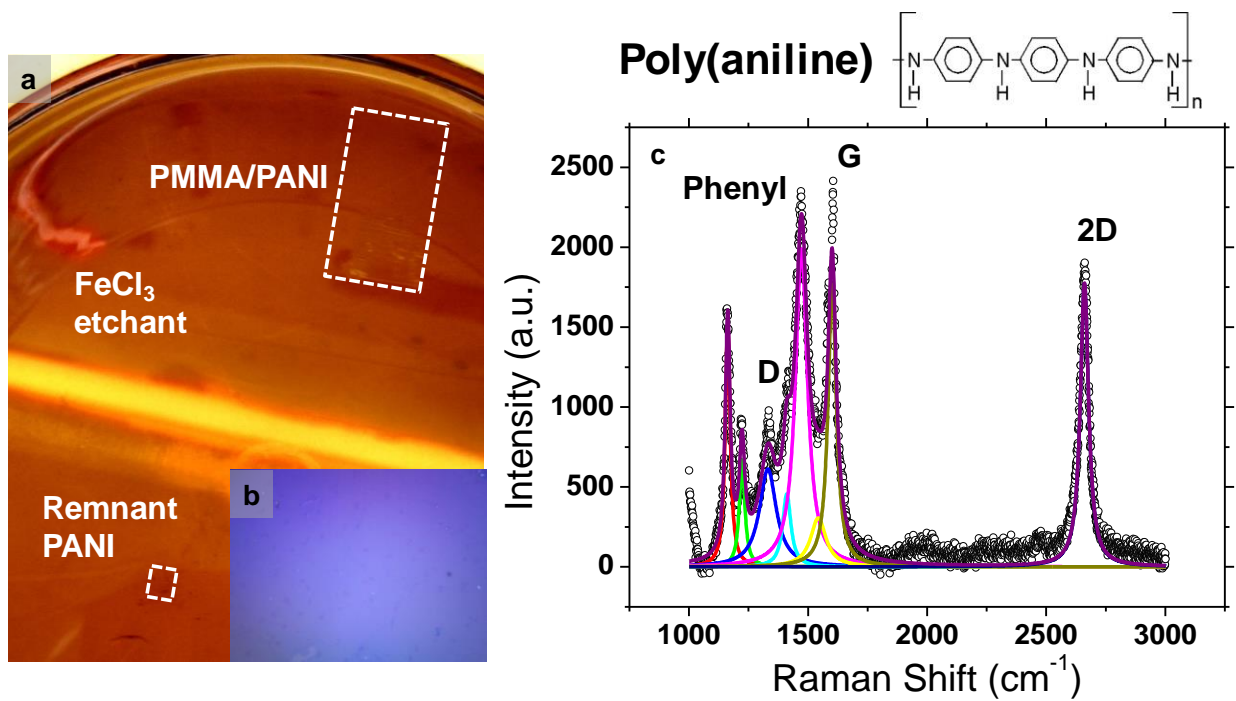

**Figure S5. Graphene transfer with aromatic poly(aniline).** **(a)** Photograph of poly(aniline) (PANI) transferred graphene and PMMA/PANI-transferred graphene on $FeCl_3$ Cu etchant. The PANI-supported graphene, without a PMMA overlayer, breaks apart within the etchant. The PMMA/PANI supported film survives the transfer. **(b)** Optical image of PMMA/PANI transferred graphene after polymer removal (in chloroform) on 90 nm $SiO_2$/Si. **(c)** Raman spectrum of the sample in **(b)**, showing considerable surface PANI residue. The spectrum shows vibrational modes related to the phenyl group[42] in PANI and as well as other hydrocarbon stretch modes. We hold that the residue level originates from strong π–π interactions between the aromatic phenyl group and the graphene.

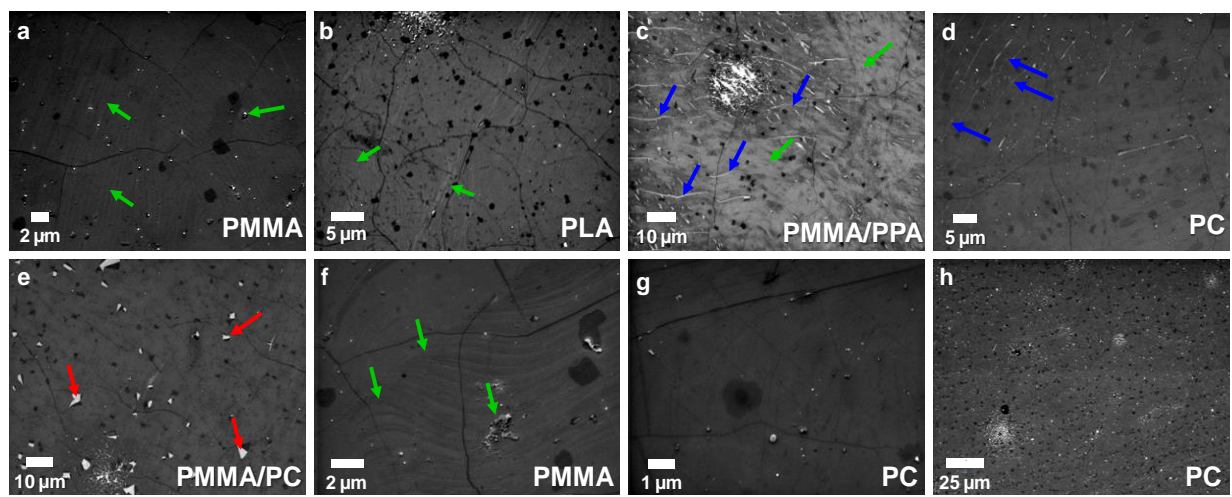

**Figure S6. Scanning electron microscopy (SEM) imaging of graphene on 90 nm SiO₂/Si, transferred by different polymers**. Green arrows show contamination, blue arrows show wrinkles, and red arrows show film breaks. Polymers removed in chloroform with no annealing. **(a)** PMMA transferred graphene with contamination. **(b)** PLA transferred graphene for a poorer quality graphene growth than (a) **(c)** PMMA/PPA bilayer transferred graphene, with low film integrity and high contamination. **(d)** PC transferred graphene for the same growth as (a). No obvious polymer contamination present. **(e)** PMMA/PC bilayer transferred graphene. PC layer contacts the graphene, with the PMMA layer providing structural support. Tears are evident, with no significant contamination. **(f)** Close-up image of the PMMA transferred sample in (a), showing larger-scale debris. **(g)** Close-up image of the PC transferred sample in (d). Only graphene bilayers, grain boundaries, and wrinkles evident. **(h)** Another PC transferred sample for a poorer quality graphene growth. Despite the growth, the polymer contamination is minimal.

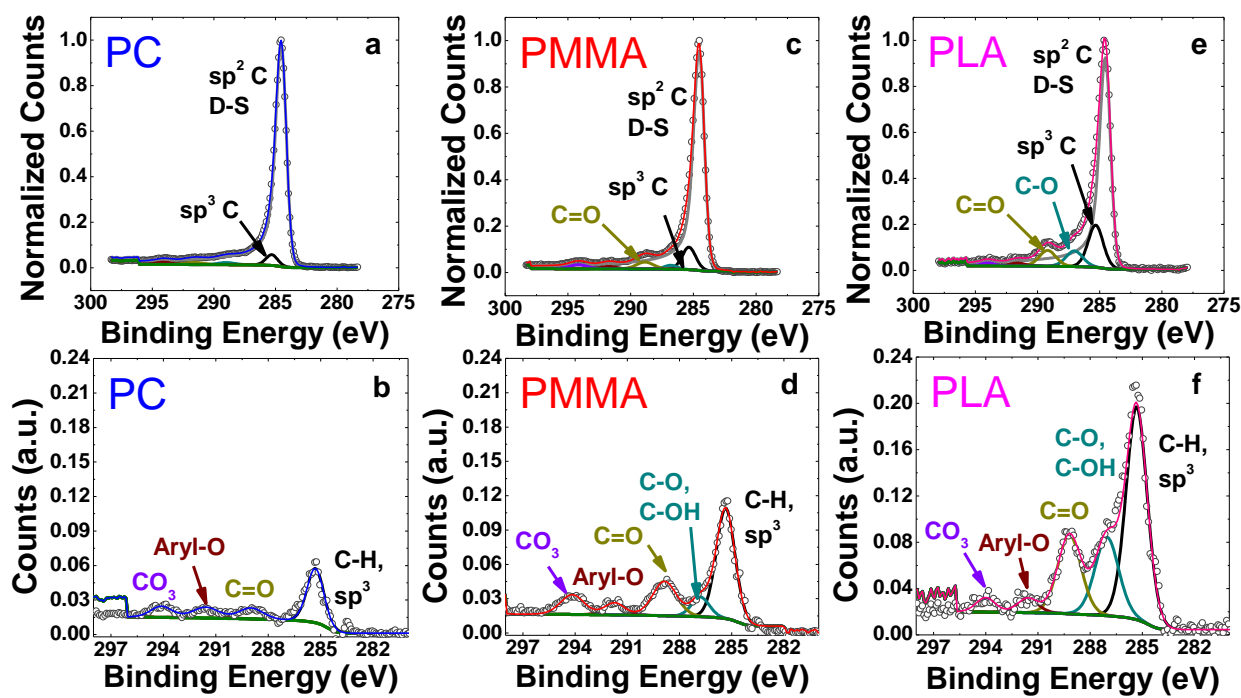

**Figure S7. X-ray photoelectron spectra (XPS) for graphene transferred with different polymers.** C 1s core level spectra (normalized to graphene) for PC-transferred (**a**), PMMA-transferred (**c**), and PLA-transferred (**e**) graphene films on 90 nm $SiO_2/Si$. A Doniach-Sunjic (D-S) lineshape was used to fit the asymmetric $sp^2$ carbons characteristic of metallic graphene. Other functionals, such as $sp^3$ carbon, carboxyls C–O, and carbonyls C=O, are also shown. PC-transferred graphene shows no obvious sub-peaks, indicative of low amounts of residue. Zoomed-in C 1s core level spectra for PC-transferred (**b**), PMMA-transferred (**d**), and PLA-transferred (**f**) graphene, with the $sp^2$ carbon contribution removed. PC-transferred graphene shows a small $sp^3$ peak, likely resulting from the graphene CVD growth process itself. Additionally, some weak carbonyl, oxygenated aryl, and carbonate ($CO_3$) groups are present. PMMA-transferred graphene has more significant contributions for the different functional groups, as compared to PC-transferred graphene. The higher $sp^3$ peak also corresponds to more aliphatic groups, like the end methoxy group and the C–C backbone in PMMA. PLA-transferred graphene is more substantially contaminated than PMMA-transferred graphene, despite attempted PLA gasification at temperatures above 180°C.[13] Large contributions from $sp^3$ carbon, carboxyls, and carbonyls originate from the aliphatic ester and ether linkages within the PLA repeat unit. To quantitatively analyze the residue differences between PMMA- and PLA-transferred graphene with respect to PC-transferred graphene, we subtract the PC $sp^3$ contribution (in area) and oxygenated aryl from the other two samples' spectra. We then sum the resultant sub-peak areas of Figs. S6b,d,f. These sums are compared relative to the D-S $sp^2$ peak area, thereby giving the PMMA, PLA, and PC percentages reported in the main manuscript.

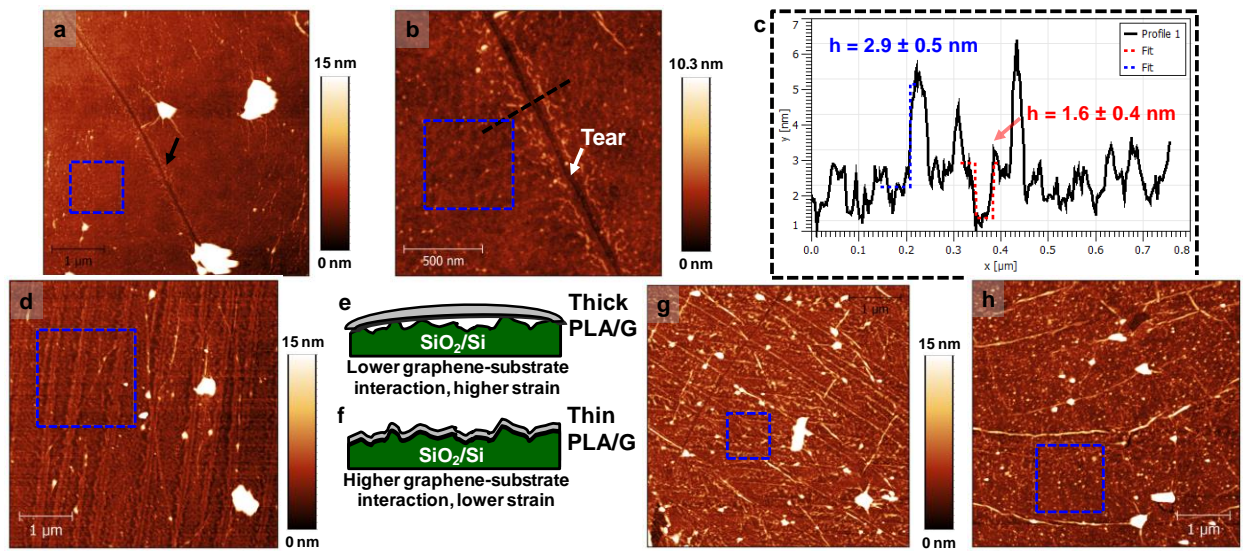

**Figure S8. Poly(lactic acid) (PLA) and poly(phthalaldehyde) (PPA) transfer of graphene. (a)** AFM image of PLA transferred graphene with a tear in the film. Sample annealed at 400°C in a low pressure (~1 torr) environment to gasify[13] the PLA. RMS roughness within the blue box is 0.63 nm, and the image's RMS roughness is 6.83 nm. **(b)** AFM close-up of the region indicated by the arrow in (a). RMS roughness within the blue box is 0.49 nm, and the entire image's RMS roughness is 0.67 nm. **(c)** Topographic profiles of the region indicated by the black line in (b). The red cut over the graphene tear reveals monolayer graphene (with water and adsorbates), whereas the blue cut shows PLA decoration near the tear. **(d)** AFM of a different PLA transfer, wherein the PLA solution for (a) and (b) was diluted in chloroform (5:1, chloroform : original PLA). This sample underwent a 200°C anneal in a low pressure (~1 torr) environment as well. RMS roughness within the blue box is 1.04 nm, and image's RMS roughness is 3.34 nm. The surface is markedly contaminated with PLA residue, despite the more dilute polymer solution used for transfer. **(e)** Cartoon schematic of a thick PLA/G film on a SiO$_2$/Si surface. Here, the thicker polymer prevents the graphene from coming into intimate contact with the SiO$_2$/Si substrate. This increases strain but lowers graphene's influence on polymer dissolution and gasification. Thus, the gasification of PLA at temperatures above 180°C proceeds as expected for the bulk polymer. **(f)** Cartoon schematic of a thin PLA/G film on a SiO$_2$/Si surface. In this case, the thin polymer allows the graphene to be conformal to the SiO$_2$/Si substrate. Consequently, this lowers the amount of strain in the graphene but increases the graphene-substrate interaction. That increased interaction affects PLA gasification[13] above 180°C. **(g)** AFM image of PPA transferred graphene employing a PMMA overlayer (PMMA/PPA bilayer, with PPA contacting the graphene). RMS roughness within the blue box is 1.45 nm, and the image's RMS roughness is 3.38 nm. Tears, wrinkles, and contamination are present. **(h)** Additional AFM image of PMMA/PPA transferred graphene. RMS roughness within the blue box is 1.24 nm, and the image's RMS roughness is 3.77 nm. Image possesses similar contamination as (g).

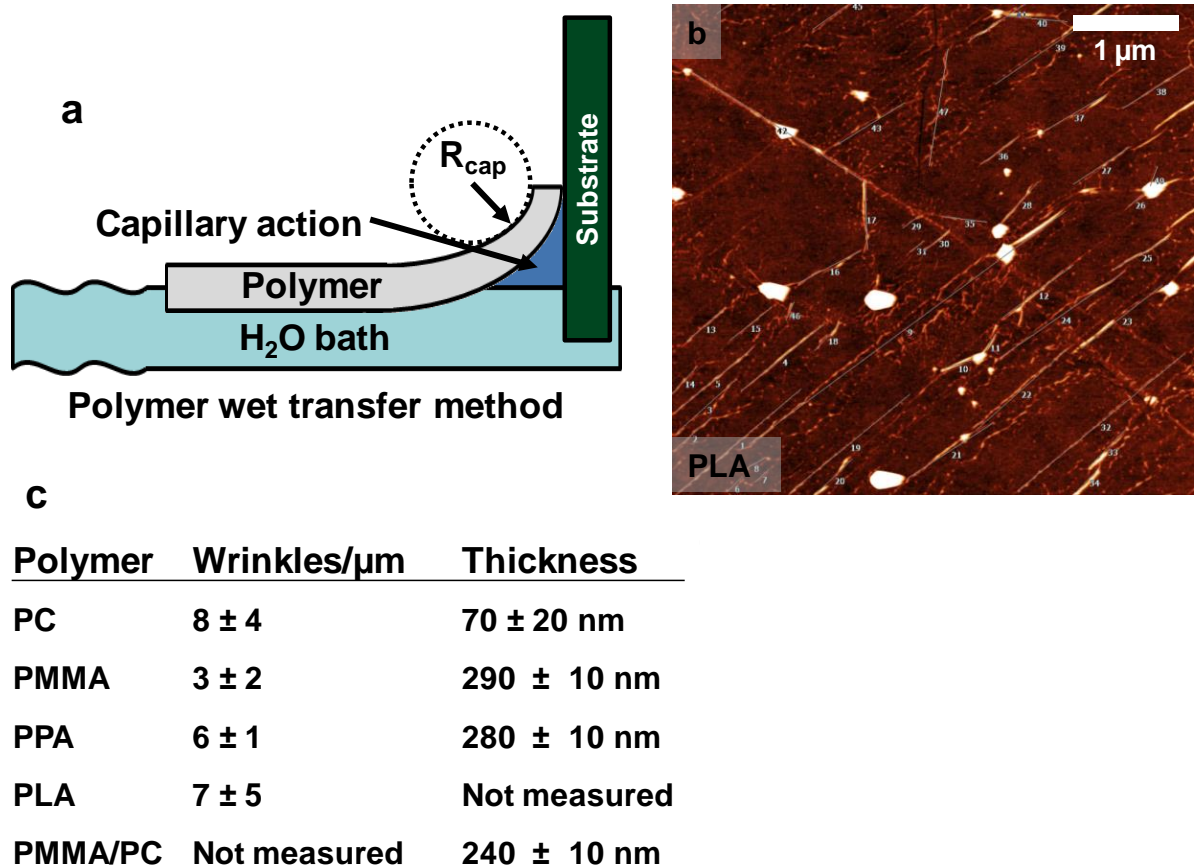

**a**

Capillary action

$R_{cap}$

Substrate

Polymer

$H_2O$ bath

**Polymer wet transfer method**

**b**

1 µm

PLA

**c**

| Polymer | Wrinkles/µm | Thickness |
|---|---|---|
| PC | 8 ± 4 | 70 ± 20 nm |
| PMMA | 3 ± 2 | 290 ± 10 nm |
| PPA | 6 ± 1 | 280 ± 10 nm |
| PLA | 7 ± 5 | Not measured |
| PMMA/PC | Not measured | 240 ± 10 nm |

**Figure S9. Wrinkling induced by the polymer transfer scaffold. (a)** Cartoon schematic of the "wicking" process by which the polymer/graphene is placed on an arbitrary substrate. This process introduces strain and mechanical wrinkling in the film. **(b)** AFM image of PLA transferred graphene, with wrinkle locations annotated in the image. **(c)** Table of average wrinkle density (per µm) for different transfer scaffolds. The scaffold thicknesses are also listed, as determined by profilometry.

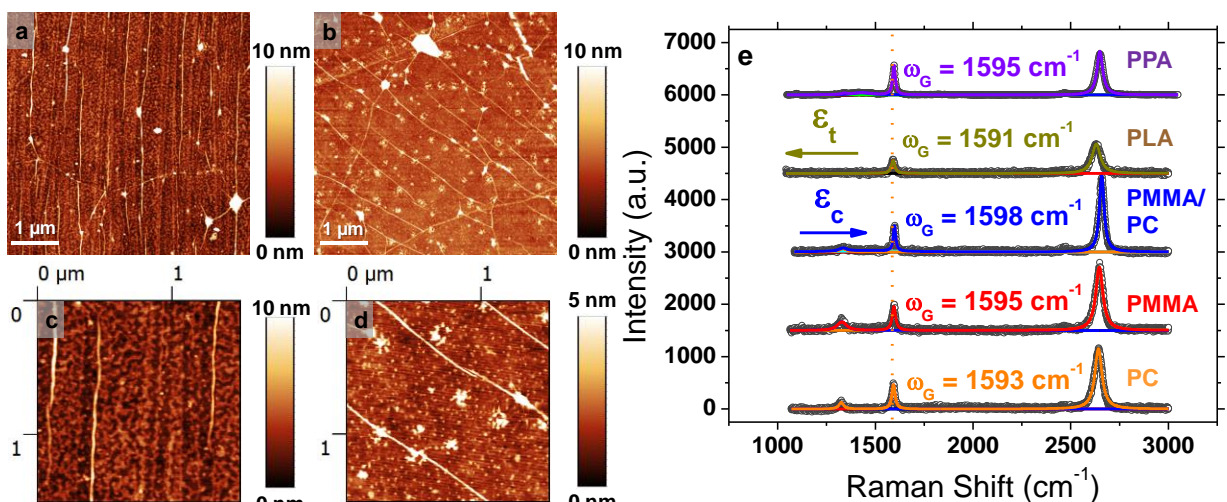

**Figure S10. Graphene doping from trapped water under graphene on SiO₂/Si.** (**a**) AFM image of PMMA-transferred graphene on SiO₂/Si with no anneal. Note that this region is reasonably free of PMMA contamination, but regions like this were rare. Tendril-like features[2] between the graphene wrinkles are evident, showing trapped water at the graphene-SiO₂/Si interface. (**b**) AFM image of PC-transferred graphene on SiO₂/Si with no anneal. Tendrils are also present, but the sample possess more point-like water features. Close-up AFM images of the PMMA (**c**) and the PC (**d**) samples in (a) and (b), respectively. Aforementioned water features are more obvious in (c) and (d). (**e**) Point Raman spectra for all of the transfer scaffolds: PPA (with a PMMA support), PLA, PMMA/PC, PMMA, and PC. The upshifted G band position ($\omega_G > 1590$ cm⁻¹) and downshifted 2D band position ($\omega_{2D} < 2655$ cm⁻¹ at $\lambda_{exc} = 633$ nm) reveal common n-type doping for all the non-annealed, transferred films. This n-type doping is induced by the trapped water (see Fig. S7). Additional upshifts in the G band result from p-type doping caused by the adsorbed polymer contamination. PC transferred films possess the lowest amount of doping, supporting the conclusion that they dissolve off the graphene top-side cleanly. Compressive strain is present in the PMMA/PC film, and tensile strain is evident in the PLA film. All Lorentzian-fitted values are given for these Raman spectra in Table S1.

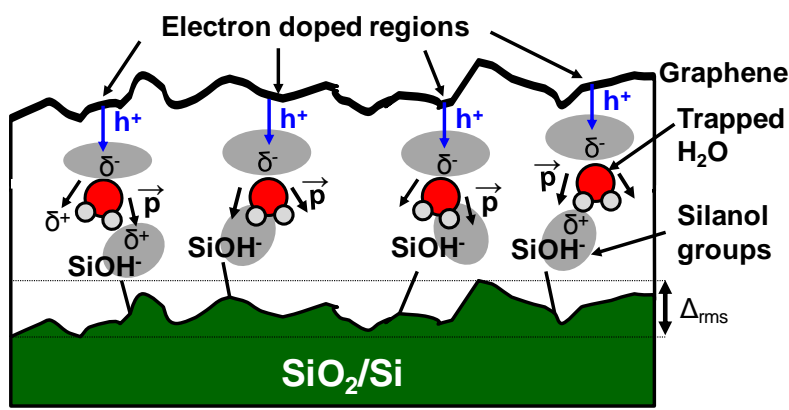

**Figure S11. Mechanism of water-induced n-type doping in graphene.** As shown in the schematic, the SiO₂/Si surface can expose silanol (Si–OH) functionals, which tend to be negatively charged. Water has an innate dipole moment **p** which will electrostatically align the hydrogens to the charged Si–OH⁻. This places the electronegative (δ⁻) oxygen into alignment with the graphene overlayer. Hole transfer to the electronegative oxygen leaves an accumulation of electrons within the graphene, thereby n-type doping the layer. The density of Si–OH groups in dry oxidized SiO₂/Si (90 nm) is $n_{imp} = 8\times10^{18}$ cm⁻³.[43] Within a 1 nm RMS

roughness ($\Delta_{rms}$) exposed layer, the estimated surface density of Si–OH groups is $n_s = 8 \times 10^{11}$ cm$^{-2}$. Assuming that multiple water molecules could be electrostatically attracted to a single Si–OH moiety, an electron concentration of $n \sim 10^{12}$ cm$^{-2}$ could be induced. This is in qualitative agreement ($n_{Raman} \approx 4 \times 10^{12}$ cm$^{-2}$) with the G band upshift and 2D band downshift observed in Fig. S10.

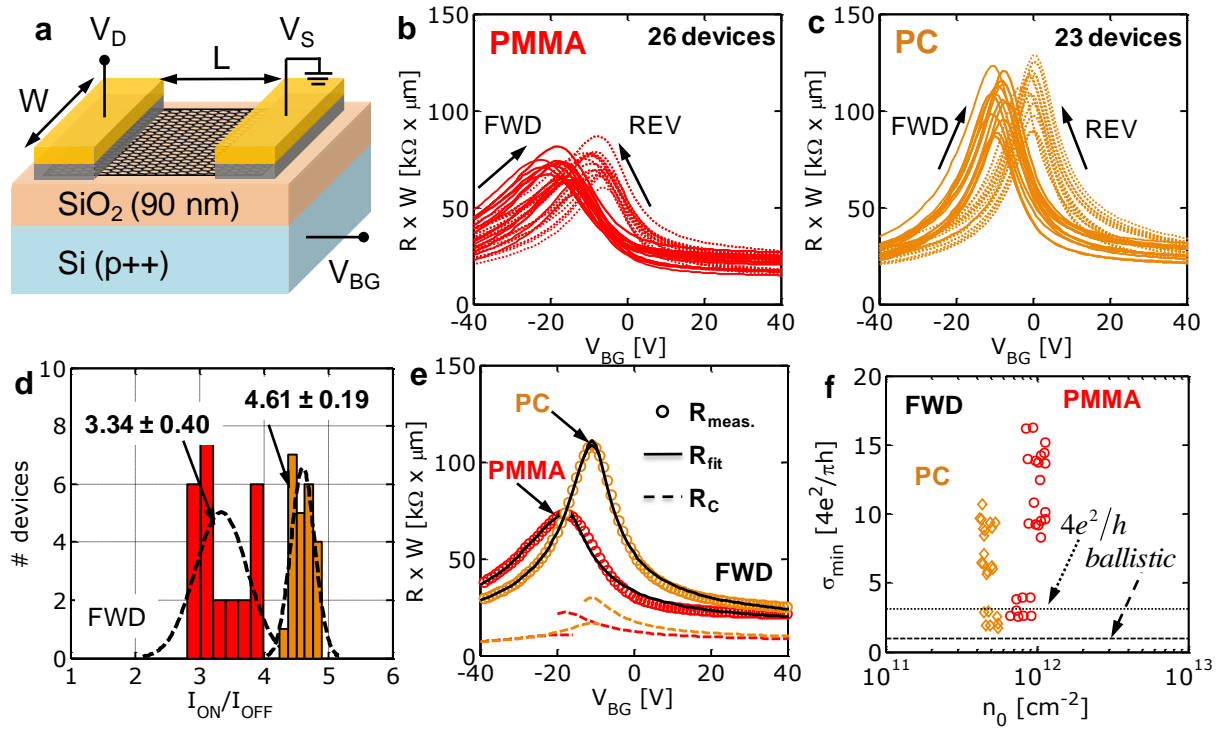

**Figure S12. Electrical characteristics of graphene field-effect transistors (GFETs) in vacuum**. Graphene films transferred using PMMA and PC based scaffolds without undergoing any thermal annealing. (**a**) Schematic of back-gated GFETs used in this work. Transfer characteristics for PMMA (**b**) and PC (**c**) based FETs respectively. Back gate voltage ($V_{BG}$) is swept consecutively in forward (FWD) and reverse (REV) directions. A shift in Dirac voltage ($\Delta V_0$) as FWD and REV sweeps are completed is observed in both cases. This n-type hysteresis suggests the presence of charge trapping mechanisms at the graphene/SiO$_2$ interface, possibly from left over polymer (PMMA or PC) residues. (**d**) $I_{ON}/I_{OFF}$ ratios for PMMA (red) and PC (orange) based FETs devices. Note that from histograms and distributions that GFETs from PC transferred films exhibit higher $I_{ON}/I_{OFF}$ and reduced device variability. (**e**) $R$-$V_{BG}$ measured (circles) and fitted (solid lines) data from Fig. 5 (b-c). Fitted electron and hole contact resistances ($R_C$) (dashed lines) also shown. Fitting model described in Ref. 11. (**f**) Measured minimum conductivity ($\sigma_{min}$) as a function of minimum carrier density ($n_0$) extracted from transport model.[11] Ballistic and $4e/h^2$ limits also shown with dashed and dotted lines, respectively.

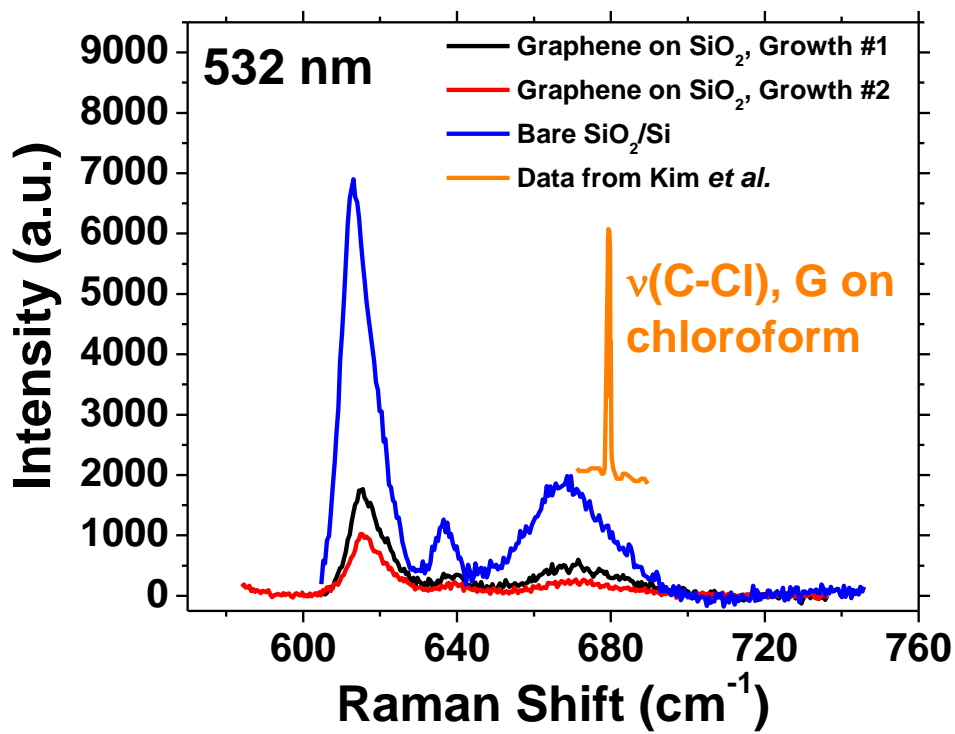

**Figure S13. Lack of chloroform intercalation under graphene.** Point Raman spectra ($\lambda_{exc}$ = 532 nm, ~10 mW power, 100X, 60 s acquisition) of two different graphene growths on $SiO_2/Si$ and a bare $SiO_2/Si$ control. Chloroform has been shown to intercalate under graphene,[44] giving the Raman signature seen in orange above. Since the polymer scaffold in our samples is often removed by chloroform, we consider the possibility that chloroform could intercalate under our graphene films. We only see signatures of the $SiO_2/Si$ in the region where intercalated chloroform modes are expected. Therefore, we conclude the amount of intercalated chloroform under the graphene is minimal.

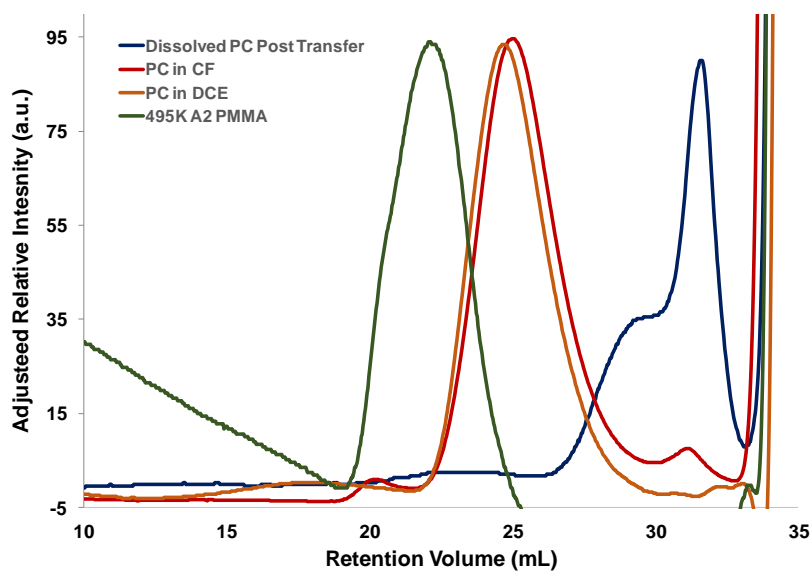

**Figure S14. Gel permeation chromatography (GPC) of PC before and after dissolution. (a)** GPC of our commercial PMMA (495K in anisole solvent, 2% by wt.), PC in CF (1.5% by wt.), PC in DCE (3% by wt.), and the dissolved PC after graphene transfer (from a PC dispersed in DCE scaffold). From analysis of the calibrated intensity versus the retained volume, we determine that the PMMA has a molecular weight (MW) of 470 kDa (polydispersity PDI of 2.7), the PC-CF has a MW of 46 kDa (PDI = 2.1), the PC-DCE has a MW of 51 kDa (PDI = 1.8), and the PC post transfer has a possible MW of 1-2 kDa. In the dissolved PC case, the GPC shows a shoulder which likely corresponds to dissolved PC oligomers (MW of 1-2 kDa). We note that the signal-to-noise ratio here is low from the small (~µg) amount of PC mass dissolved during transfer. Thus, this shoulder could occur from instrument noise and/or impurities in the tetrahydrofuran (THF) solvent. Void volume peak is at a retention volume of 34.8 mL for all polymers.

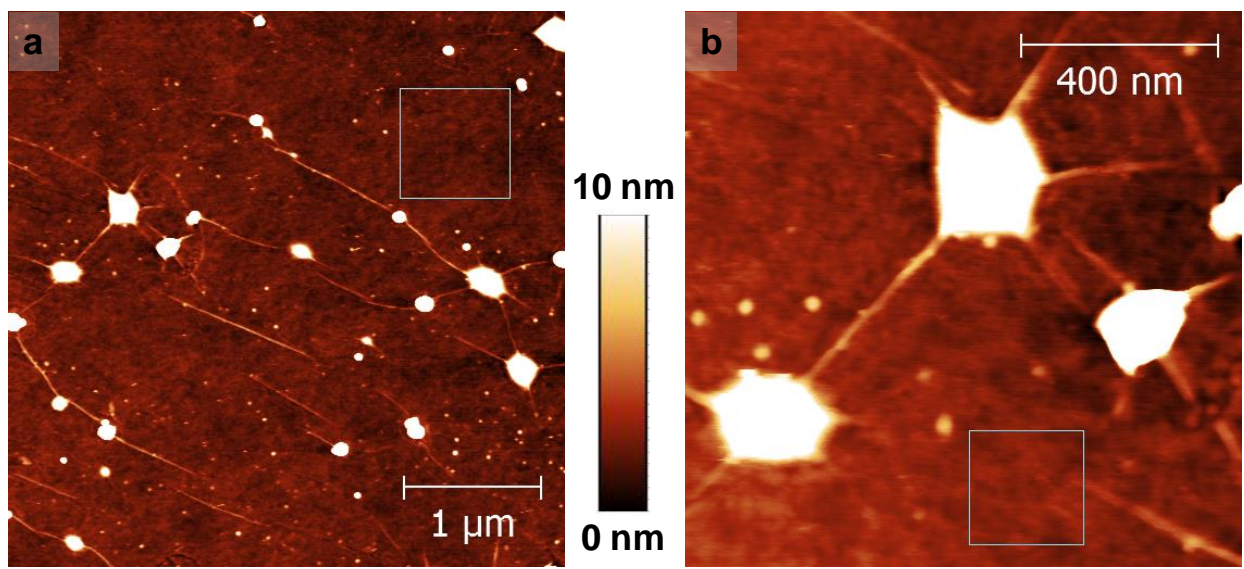

**Figure S15. Clean graphene transfer with low molecular weight PMMA. (a)** Large-area AFM image of graphene transferred with a 4K molecular weight (MW) PMMA layer supported by a 495K overlayer. RMS roughness is 0.35 nm in the box, 5.11 nm in the image. Both layers were spun at 3000 RPM for 30 s

and baked out at 200°C for 2 min. A PMMA overlayer was necessary, as unsupported 4K PMMA scaffolds broke apart mechanically in the water rinses. Low MW polymers have a smaller footprint on the graphene. This smaller footprint circumvents possible interaction with graphene morphologies like wrinkles and grain boundaries. Consequently, this promotes effective polymer dissolution if the polymer does not electrostatically interact with the graphene, as is the case with aliphatic PMMA. **(b)** Small-area AFM image of graphene transferred with 4K PMMA, showing again a smooth film with trapped water at the large protrusions. RMS roughness is 0.31 nm in the box, 3.77 nm in the image.

| Sample | D band frequency (cm⁻¹) | D band FWHM (cm⁻¹) | G band frequency (cm⁻¹) | G band FWHM (cm⁻¹) | 2D band frequency (cm⁻¹) | 2D band FWHM (cm⁻¹) | I(2D) /I(G) | A(D) /A(G) | Strain |
|---|---|---|---|---|---|---|---|---|---|
| **PC** | 1325.5 | 28.0 | 1592.7 | 19.7 | 2641.8 | 43.5 | 2.50 | 0.39 | N/A |
| **PMMA** | 1328.3 | 56.8 | 1594.8 | 21.0 | 2646.6 | 44.0 | 2.65 | 0.97 | N/A |
| **PMMA/ PC** | 1334.4 | 50.1 | 1598.0 | 15.3 | 2659.0 | 28.9 | 3.00 | 0.41 | −0.55% (compressive) |
| **PLA** | N/A | N/A | 1590.9 | 21.4 | 2630.7 | 50.6 | 2.27 | N/A | +0.73% (tensile) |
| **PPA** | 1340.6 | 38.6 | 1595.3 | 16.6 | 2649.4 | 34.8 | 1.48 | 0.09 | N/A |

**Table S1. Raman metrics from point spectra for different polymer scaffolds.** Note that the peak heights are used for I(2D)/I(G) ratio, whereas the area values are used for the A(D)/A(G) ratio. For the strain calculations, we calculated from a G band position[32] of 1584 cm⁻¹ and employed previously reported Grüneisen parameters and strain-based shifts.[33] We assumed the water's n-type contribution (Fig. S9) gave a 9 cm⁻¹ upshift of the G band from the intrinsic G band position of graphene (1584 cm⁻¹) and the G band position of the PC transferred graphene (1593 cm⁻¹). The PLA transferred film has an additional n-type shift of ~4 cm⁻¹ from the adsorbed residue.

| Polymer (on SiO$_2$/Si) | Thickness (nm) | Spin Conditions | Bakeout Conditions |
|---|---|---|---|
| **495K PMMA, then 950K PMMA** | 290 ± 10 nm | 3000 RPM, 30 s for 495K; 3000 RPM, 30 s for 950K | 200°C for 2 min for each PMMA layer |
| **495K PMMA, then 950K PMMA** | 235 ± 15 nm | 3000 RPM, 30 s for 495K; 3000 RPM, 30 s for 950K | None |
| **4K PMMA (2% wt. in anisole)** | 23 ± 2 nm | 3000 RPM, 30 s | 200°C for 2 min |
| **4K PMMA** | 24 ± 2 nm | 3000 RPM, 30 s | None |
| **4K PMMA, then 495K PMMA** | 59 ± 4 nm | 3000 RPM, 30 s | 200°C for 2 min for each PMMA layer |
| **PC dispersed in chloroform (CF)** | 70 ± 20 nm | 3000 RPM, 30 s | None |
| **PC dispersed in CF** | 80 ± 20 nm | 3000 RPM, 60 s | None |
| **PC dispersed in CF** | 60 ± 15 nm | 5000 RPM, 30 s | None |
| **PC dispersed in CF** | 60 ± 15 nm | 7000 RPM, 30 s | None |
| **PC dispersed in CF, then 495K PMMA, then 950K PMMA** | 240 ± 20 nm | 3000 RPM, 30 s for PC; 3000 RPM, 30 s for 495K; 3000 RPM, 30 s for 950K | None for PC layer; 200°C for 2 min for each PMMA layer |
| **PC dispersed in CF, then 495K PMMA, then 950K PMMA** | 295 ± 10 nm | 3000 RPM, 30 s for PC; 3000 RPM, 30 s for 495K; 3000 RPM, 30 s for 950K | None |
| **PC dispersed in di-chloroethane (DCE)** | 40 ± 3 nm | 3000 RPM, 30 s | None |
| **4K PMMA, then PC dispersed in DCE** | 45 ± 5 nm | 3000 RPM, 30 s | None |
| **0.24 g PPA dispersed in 18 mL CF** | 60 ± 5 nm | 3000 RPM, 30 s | None |
| **0.24 g PPA solution** | < 10 nm − not reliable | 3000 RPM, 30 s | 200°C for 2 min |
| **0.16 g PPA dispersed in 15 mL CF, then 495K PMMA, then 950K PMMA** | 230 ± 10 nm | 3000 RPM, 30 s for PPA; 3000 RPM, 30 s for 495K; 3000 RPM, 30 s for 950K | None |
| **0.16 g PPA solution, then 495K PMMA, then 950K PMMA** | 280 ± 10 nm | 3000 RPM, 30 s for PPA; 3000 RPM, 30 s for 495K; 3000 RPM, 30 s for 950K | None for PPA layer; 200°C for 2 min for each PMMA layer |
| **60K phenyl methacry-late in CF** | < 20 nm *ca.* 10 ± 5 nm | 3000 RPM, 30 s | 200°C for 2 min |
| **60K phenyl methacry-late in CF** | < 20 nm *ca.* 10 ± 5 nm | 3000 RPM, 30 s | None |
| **Poly(aniline) (PANI) in CF** | < 10 nm − not reliable | 3000 RPM, 30 s | 200°C for 2 min |
| **PANI in CF, then 495K PMMA** | 36 ± 6 nm | 3000 RPM, 30 s | 200°C for 2 min for each layer |

**Table S2. Thickness of different polymer scaffolds.** All polymers are placed on 90 nm SiO$_2$/Si witnesses and not graphene on Cu substrates, and the thicknesses are determined by profilometry.